\documentclass[11pt, english, aps, tightenlines, prd, nofootinbib]{revtex4-2}
\usepackage[utf8]{inputenc}
\usepackage{babel}
\usepackage[dvipsnames]{xcolor}
\usepackage{array}
\usepackage{multirow}
\usepackage{amsmath,amssymb}
\usepackage{dsfont}
\usepackage{cancel}
\usepackage{stackrel}
\usepackage{braket}
\usepackage{physics}
\usepackage{fontawesome}
\usepackage{graphicx}
\usepackage{subcaption}
\graphicspath{{./}}

\usepackage{breakcites}
\usepackage{natbib}
\usepackage{geometry}
\usepackage[unicode=true,pdfusetitle,
 bookmarks=true,bookmarksnumbered=false,bookmarksopen=true,
 breaklinks=false,backref=false,colorlinks=false]
 {hyperref}
\renewcommand{\d}{\mathrm{d}}
\newcommand{\iu}{\mathrm{i}}
\newcommand{\ee}{\mathrm{e}}
\renewcommand{\Im}{\mathfrak{Im}}

\begin{document}

\title{Quantum and Gradient Corrections to False Vacuum Decay on a de Sitter Background}

\author{Juan S. Cruz}
\affiliation{CP3-Origins, Center for Cosmology and Particle Physics Phenomenology, University of Southern Denmark, Campusvej 55, 5230 Odense M, Denmark}
\email{jcr@sdu.dk}

\author{Stephan Brandt}
\email{stephan.brandt@physik.uni-muenchen.de}
\affiliation{Ludwig-Maximilians Universität\\
	Theresienstr. 37, 80333 München, Germany}
\author{Maximilian Urban}
\email{maximilian.urban@physik.uni-muenchen.de}
\affiliation{Ludwig-Maximilians Universität\\
	Theresienstr. 37, 80333 München, Germany}

\date{August 10, 2022}

\begin{abstract}
We study the effects of a fixed de Sitter geometry background in scenarios of false vacuum decay. It is currently understood that bubble nucleation processes associated with first order phase transitions are particularly important in cosmology. Considering the geometry of spacetime complicates the interpretation of the decay rate of a metastable vacuum. However, the effects of curvature can still be studied in the particular case where backreaction is neglected. We compute the imaginary part of the action in de Sitter space, including the one-loop and the gradient corrections. We use two independent methodologies and quantify the size of the corrections without any assumptions on the thickness of the wall of the scalar background configuration.
\end{abstract}

\maketitle

\section{Introduction}
Cosmological phase transitions have been interesting historically for different reasons and remain an active field of study. The current state of the Standard Model (SM) of particle physics is that, given the masses of the top quark and the Higgs boson, the effective potential at at Next-to-nextto-leading order (NNLO) approximation displays two minima, one of which is associated with a lower vacuum energy appearing at high field values, explored probably at extremely high energies $\sim 10^{11}$ GeV \cite{Degrassi:2012ry, Buttazzo:2013uya}. This means that at zero temperature, the effective SM potential presents us with a metastable situation. Although it is known that the Electroweak Phase Transition (EWPT) is not of first-order \cite{Kajantie:1995kf} when including the effects of temperature, such a process remains interesting to study in the context of cosmology, especially in possible extensions to the SM since this type of phase transitions is present in a variety of different contexts. One example are early models of inflation \cite{Guth:1980zm, Turner:1992tz}, which feature a first order phase transition. It also appears in Baryogenesis, which addresses the matter and anti-matter asymmetry or Baryon asymmetry of the universe (BAU)  \cite{Kuzmin:1985Anomalous}, where some of the models rely on strong enough first-order EWPTs. An even more interesting relation might even be that of the Higgs and inflation as in Higgs-inflation \cite{Bezrukov:2007ep}. This model would bring these different ideas together, making the Higgs field responsible for both phenomena.

Alternatively, a remarkable feature of phase transitions in the early universe is the generation of gravitational waves, as has been suggested \cite{Hogan:1986Gravitational, Witten:1984rs}. While this is still a topic of discussion (see Ref.~ \cite{Hindmarsh:2020hop} for a review), current technology is reaching the point where we may observe gravitational waves from inflation \cite{Bartolo:2016ami, Caldwell:2022qsj}, so that an understanding of how often bubbles of true vacuum can be nucleated may lead to a better insight into the details of a possible early phase transition. Eventually, this will serve to check possible signals in the Stochastic Gravitational Waves Background (SGWB)  \cite{Hindmarsh:2014Gravitational}. Other studies have attempted to put bounds on the Higgs coupling to the curvature scalar by studying similar settings during inflation \cite{Mantziris2020} or to model the exit of inflation via a first order phase transition  \cite{Adams:1990ds}. For all of these applications, it is important to study in detail how first order phase transitions occur in field theory on non-flat backgrounds.

First order phase transitions are characterized by a tunneling process running through a potential barrier between local minima. In general, the minimum sitting at higher energy is called the false vacuum, while we refer to the lowest-lying one as the true vacuum. The path integral formalism allows us not only to approximate the decay rate of the false vacua at $0^{\text th}$ order but also to expand the action around inhomogeneous solutions and to consider the contributions coming from quantum fluctuations. After these ideas were initially put forward by Callan and Coleman \cite{Coleman1977},  \cite{Callan1977}, there have been several studies around false vacuum decays in different settings, some of which attempt to introduce gravitational effects through the consideration of more general geometries \cite{Coleman1980,Hawking1983}.

In this communication, we compute the decay rate of the false vacua in a theory with a scalar field in a fixed de Sitter background, subject to a potential allowing for two distinct local minima. De Sitter space is particularly important in accelerated expansionary phases of the universe, such as inflation or the current epoch. Therefore, studying how the decay rate of the false vacuum may be influenced by curved spacetimes, or in particular, a cosmological constant, can lead to a better understanding of these stages (see e.g.~ \cite{Joti:2017fwe}). A review of the metastability of the vacuum is given in  \cite{Markkanen2018}.

We present two methods that permit the computation of contributions of higher order effects in a fixed de Sitter space background. First, we compute the associated functional determinants, using the Gel'fand-Yaglom theorem \cite{Gelfand:1959nq} and Green's functions \cite{Baacke:1993jr}. Later we include the backreaction of gradient effects fed back into the scalar field background, expanding on ideas first discussed in~ \cite{Dunne2005, Hackworth:2004xb,Garbrecht2015a}. We include one-loop effects with gradients in a zero temperature setting, which is related to studies such as  \cite{Salvio:2016mvj, Rajantie2018ajw, Camargo-Molina:2022paw}. We focus specifically on computing the imaginary part of an effective action for bounce-like field configurations.

Traditional treatments of the computation of the decay rate rely on a number of approximations, such as ignoring the gradients of the field or assuming degenerate vacua. In this paper, we focus on computing the effects of a fixed geometrical background, that is we neglect the dynamics of the gravitational sector. We consider a general type of wall, as opposed to a thin wall case. Self-consistent methods will be used to obtain corrections coming from the gradients of the bounce solution $\phi^{(0)}$.

The paper is organized as follows. In section \ref{sec:vacuumDecayFlat} we briefly review the theory behind vacuum decay in flat spacetime, introducing notation and concepts that will be used throughout the document. In section \ref{sec:vaccumDecayCurvedSpacetime} we describe how the expressions are adapted to a curved spacetime setting and introduce the necessary details pertaining to the de Sitter spacetime. Subsequently, in section \ref{sec:funcDets}, we compute the fluctuation operator and elaborate on the two methods we employ to calculate the functional determinants therein. In section \ref{sec:renorm} we take care of renormalizing the theory by employing the WKB expansion to obtain the divergences appearing at the one-loop level. Section \ref{sec:higherOrder} contains the computation of the tadpole functions and quantum corrections to the scalar background. We show how the previously exposed framework can be used in section \ref{sec:numericalImpl}, where the numerical treatment of a benchmark set of parameters is fully explained. We summarize our work and findings in section \ref{sec:conclusions}.

\section{Vacuum decay in flat spacetime}
\label{sec:vacuumDecayFlat}

Mainstream computations of amplitudes of physical processes generally employ the path integral formulation of quantum field theory expanded around homogeneous expectation values \cite{Jackiw:1974cv, Coleman:1973jx}. The partition function is
\begin{equation}
\langle \Omega_{\text{out}} | U(t_f,t_i) | \Omega_{\text{in}} \rangle = Z[0]\equiv \int_{\Omega_{\text{in}}}^{\Omega_{\text{out}}} \mathcal{D}[\phi] \ee^{\frac{\iu}{\hbar} S[\phi]},
\label{eq:vacuumEvolution}
\end{equation}
which corresponds to the transition between the vacua $\Omega_{in}$ and $\Omega_{out}$.   These appear as boundary conditions on the RHS, where it is understood that the functional integration covers all field configurations that satisfy $\phi(t_i,\vec{x})=\Omega_{\text{in}}(\vec{x})$ and $\phi(t_f,\vec{x})=\Omega_{\text{out}}(\vec{x})$.

\begin{figure}[htbp]
	\includegraphics[clip,width=.5\textwidth]{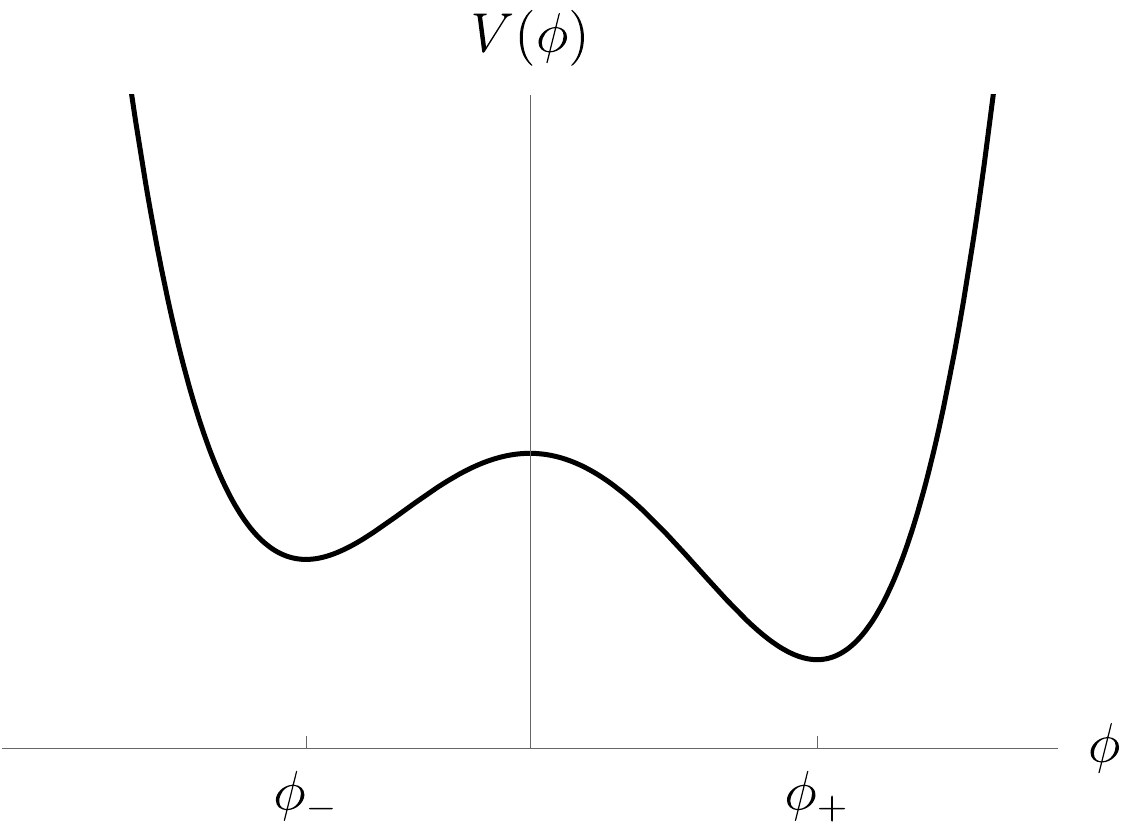}
	\caption{Example of a scalar potential displaying several minima which are non-degenerate. The higher local minimum is referred to as the false vacuum, while the right local minimum which stands at a lower potential is called the true vacuum.}
	\label{fig:treeLevelPotential}
\end{figure}

Let us then consider a field theory for a single real scalar field $\phi(x)$, subject to a potential $V(\phi(x))$, which is analytic and displays more than one local minimum, as seen in Fig.~\ref{fig:treeLevelPotential}. The standard Lagrangian density for such a theory is
\begin{equation}
\mathcal{L} = \frac{1}{2}(\partial_x \phi(x))^2 - V(\phi(x)),
\end{equation}
where $V(\phi(x))$ is such that there exist at least two solutions to $\frac{\partial}{\partial \phi}V(\phi(x)) = 0$ associated with two minima, say for field values $\phi_{\pm}$.
This feature of the potential implies the existence of an additional saddle point in the path integral, meaning an overlap of states that are located at each minima. As we will see, there exist field configurations that interpolate between these two field values, for which the action is finite. Generically, one of these field values will be associated with a higher potential; take $\phi_- > \phi_-$ and $V(\phi_-)\leq V(\phi_-)$ and add a constant to the potential as to have $V(\phi_-)=0$. In this setup, a static field configuration $\phi(t,\vec{x})=\phi_-$ will a have a zero action, while a field configuration located at the other minimum, $\phi(t,\vec{x})=\phi_-$, will have a negative action. As both are saddle points of the theory, one can expand the action around a scalar field having an expectation value about either of these vacua. These two settings are usually referred to as true and false vacuum, respectively, where the \emph{false} one corresponds to the minimum associated with a higher potential.

We are interested in tunneling phenomena between these two configurations. The corresponding expansion is not made around homogeneous expectation values but around a field configuration that spends an infinitely long time around each of the minima. This solitonic configuration, known as the bounce, can be shown to correspond to imaginary energy values  \cite{Coleman1977},  \cite{Coleman1985}. Therefore, the transition amplitude can be interpreted as a decay rate from the false vacuum to the true vacuum.

To compute the transition from one vacuum state to the other, we will need to employ a Wick rotation and compute in the limit where time is taken to infinity:
\begin{align}
\langle \phi_- | \ee^{\iu H T} | \phi_-\rangle =
\int_{\phi(-T)=\phi_-}^{\phi(T)=\phi_-} \mathcal{D}[\phi]
\exp\left(\frac{\iu}{\hbar}\int_{-T}^{T} \dd t\; \dd[3]\vec{x}\; \mathcal{L}\right),
\end{align}
here $T$ is a real parameter that can be analytically continued to the complex plane. We do this in two steps. First substitute $T\rightarrow \iu\mathcal{T}$, with $\mathcal{T}\in\mathbb{R}$ and obtain the following expression:
\begin{align}
\langle \phi_- | \ee^{- H\mathcal{T}} | \phi_-\rangle = \int_{\phi(-\iu
\mathcal{T})=\phi_-}^{\phi(\iu\mathcal{T})=\phi_-} \mathcal{D}[\phi]
\exp\left(\frac{\iu}{\hbar}\int_{-\iu\mathcal{T}}^{\iu\mathcal{T}} \dd t\;
\dd[3]\vec{x}\; \mathcal{L}\right).
\end{align}
The next step is to perform the analytic continuation of the time variable within the integration, $t\rightarrow i\tau$, again with $\tau\in\mathbb{R}$, which leads to:
\begin{equation}
\langle \phi_- | \ee^{- H\mathcal{T}} | \phi_-\rangle = \int_{\phi(-\iu
\mathcal{T})=\phi_-}^{\phi(\iu\mathcal{T})=\phi_-} \mathcal{D}[\phi]
\exp\left(-\frac{1}{\hbar}\int_{-\mathcal{T}}^{\mathcal{T}} \dd \tau\;
\dd[3]\vec{x}\; \Delta^4 \phi(x) + V(\phi(x)) \right).
\label{eq:flatRotatedAction}
\end{equation}
The simplest approximation to the path integral is obtained by a saddle-point expansion. Let $\phi^{(0)}$ be a saddle-point configuration, satisfying the boundary conditions of the path integral above. Then,
\begin{align}
	\langle \phi_- | \ee^{- H\mathcal{T}} | \phi_-\rangle \sim \exp\left(
	-\frac{S_E\left[\phi^{(0)}\right]}{\hbar} \right)  \int_{\phi^{(1)}(-\iu
	\mathcal{T})=0}^{\phi^{(1)}(\iu\mathcal{T})=0} \mathcal{D}[\phi^{(1)}]
	\exp\left( -\frac{1}{\hbar}\frac{\delta^2 S_E\left[ \phi^{(0)}\right]
	}{\delta(\phi^{(1)})^2} \right) \equiv K \ee^{-S_0/\hbar},
\end{align}
where the subscript $E$ denotes the use of Euclidean space. The original field was decomposed as $\phi = \phi^{(0)} + \phi^{(1)}$ and $K$ is defined as the result of the Gaussian path integral over the fluctuation operator and is known to be imaginary due to its negative eigenvalue. The negative eigenvalue is essential for the interpretation as a decay rate  \cite{Callan1977}. The formal expression on the left hand side of the equation above, can be written in terms of the energies of the eigenstates of the theory
\begin{equation}
\langle \phi_- | \ee^{-H\mathcal{T}}|\phi_-\rangle = \sum_n \langle \phi_- | n\rangle \ee^{-E_n \mathcal{T}}\langle n|\phi_-\rangle \underset{\mathcal{T}\rightarrow \infty}{\sim } \ee^{- E_0\mathcal{T}}  \langle\phi_- | \phi_+ \rangle \langle \phi_+ |\phi_-\rangle.
\end{equation}
Equating the two expressions above, it is usually argued  \cite{Coleman1985} that the decay rate per unit volume is
\begin{align}
	\frac{\Gamma_{\rm decay}}{V} = 2\lim_{V,\mathcal{T}\rightarrow \infty}\frac{|\Im Z[0]|}{V\mathcal{T}} = 2\lim_{V,\mathcal{T}\rightarrow \infty}|K|\frac{\ee^{-S_0/\hbar}}{V\mathcal{T}}.
	\label{eq:decayLeading}
\end{align}
To include higher order corrections from one-loop contributions and background gradients, one can employ the effective action construction  \cite{Jackiw:1974cv}, combined with a semi-classical expansion in $\hbar$ and the method of constrained sources  \cite{Garbrecht:2016Cte,Plascencia:2016Cgd}.

We are interested in calculating the vacuum-to-vacuum transition rate between non-trivial field configurations, as this is related to tunneling phenomena via the nucleation of true vacuum bubbles  \cite{Callan1977,Coleman1977}. We compute both one-loop and gradient effects, by considering the one-loop effective action as in previous studies \cite{Baacke2004Olc}, similarly to the usage of the Coleman-Weinberg effective potential \cite{Coleman:1973jx}. In the same context of a semi-classical expansion, a one-loop self-consistently quantum corrected evaluation of the action is \cite{Garbrecht2015a}
\begin{align}
	\Gamma^{(1)}[\phi^{(0+1)}] = S[\phi^{(0+1)}] + \frac{\hbar}{2} \log \frac{\det \mathcal{G}^{-1}(\phi^{(0+1)})}{\det \mathcal{G}^{-1}(\phi_-)},
	\label{eq:effectiveAction}
\end{align}
where quantum corrections to $\phi$, $\phi^{(1)}$, are taken into account. Above, $\mathcal{G}^{-1}$ stands for the fluctuation operator and $\phi^{(0+1)}$ is the bounce together with its quantum corrections.
 The corrected version of Eq.~\eqref{eq:decayLeading} takes on almost the same form:
 \begin{align}
 	\frac{\widetilde{\gamma}}{V} = 2\lim_{V,\mathcal{T}\rightarrow \infty}\frac{\ee^{-\Gamma^{(1)}[\phi^{(0+1)}]/\hbar}}{V\mathcal{T}}.
 	\label{eq:decayRateGradients}
 \end{align}
 We see, that the original action and the factor $K$ are replaced by the effective action in Eq.~\eqref{eq:effectiveAction}, as explained in  \cite{Garbrecht:2016Cte, Plascencia:2016Cgd}.

Our objective is to first estimate the one-loop homogeneous contributions, appearing in the last term of the effective action above, when evaluated on the bounce solution. Later we compute the full effective action, by including contributions coming from its evaluation on the quantum-corrected bounce. To compute $\Gamma^{(1)}$ in Eq.~\eqref{eq:effectiveAction} we may separate the background from the quantum corrections, $\phi^{(0+1)} = \phi^{(0)} + \hbar\phi^{(1)}$, and decompose the different contributions as follows
\begin{align}
	\Gamma^{(1)}[\phi^{(0+1)}] = S_{\rm E}[\phi^{(0)}] + B^{(1)} + B^{(2)} + \mathcal{O}(\phi^{(1)^2}),
	\label{eq:effectiveActionDecomp}
\end{align}
where
\begin{align}
	B^{(1)} &\equiv \frac{\hbar}{2} \log \frac{\det \mathcal{G}^{-1}(\phi^{(0)})}{\det \mathcal{G}^{-1}(\phi_-)},\\
	B^{(2)} &\equiv \frac{\hbar^2}{2}\int \phi^{(1)}(x)\frac{\delta}{\delta\phi(x)}\log \frac{\det \mathcal{G}^{-1}(\phi)}{\det \mathcal{G}^{-1}(\phi_-)}\bigg|_{\phi=\phi^{(0)}}.
\end{align}
The computation is self-consistent in the sense that the quantum corrections are to be obtained from the equations of motion after the one-loop contributions have been computed, and afterward fed back in into the action. We will obtain analogous quantities for the case of a curved spacetime, fixed to be de Sitter, while ignoring any backreaction on the geometry. The formulae above require modification due to the possible presence of negative or zero modes, which we will discuss later in subsection \ref{subsec:zeromodes}.

\section{Vacuum decay in curved spacetime}
\label{sec:vaccumDecayCurvedSpacetime}

Solving the vacuum Einstein field equations in four dimensions with a positive cosmological constant gives the so-called de Sitter spacetime \cite{Hawking:1973uf}. The symmetry group of de Sitter space is $SO(1,4)$, which has dimension ten in analogy to the Poincaré group for Minkowski space. It is possible to choose coordinates covering half the spacetime, namely $(t,\vec{x})\in\mathbb{R}^4$, which result in a line element in an FLRW form,
\begin{align}
	\d s^2= \d t^2 - \ee^{2H t}\d\vec{x}^2,
\end{align}
where $H$ is related to the expansion rate and the cosmological constant.
It is more propitious to choose a global coordinate patch and to use the Euclidean version of de Sitter space, as required by the saddle-point expansion shown in Sec.~\ref{sec:vacuumDecayFlat}. This is done by changing coordinates, such that the metric takes on the following form
\begin{equation}
	\d s^2=\d\tau^2+ \frac{1}{H^2}\sin^2(H \tau)\d\Omega_3^2,
\end{equation}
where $\tau\in[0,\pi/H]$ and $H$ is the Hubble constant. We can recognize a Euclidean scale factor
\begin{equation}
	a(\tau) = \frac{1}{H}\sin(H\tau),
	\label{eq:scaleFactordS}
\end{equation}
as well as compute the spacetime volume element, to find
\begin{align}
\sqrt{g_{\rm E}}\,\d^4x = a^3(\tau)\d\tau\d\Omega_3 = \frac{1}{H^3}\sin^3(H\tau)\sin^2(\theta_1)\sin(\theta_2) \d\tau\d\theta_1\d\theta_2\d\phi.
\label{eq:volumeElement}
\end{align}

\subsection{Our model}

In our approach we neglect the evolution of the scale factor according to the Friedmann equations and assume that the scalar field $\phi$ has an energy scale such that its evolution will not have an impact on the de Sitter background.

We consider a standard curved spacetime Lagrangian for a scalar field subject to a potential $V$. $V$ exhibits two different local minima and therefore allows for tunneling phenomena. We consider cases where the scale of possible bubbles is much smaller than the expansion scale, thus we fix the metric to that of de Sitter and ignore its dynamics. As in \cite{Rubakov:1999ir, Hackworth:2004xb}, this corresponds to a scenario where the background metric can be approximated by de Sitter and where simultaneously gravitational effects may not be negligible. The former is justified as long as the fractional variation of the potential is small when the field varies between $\phi_-$ and $\phi_+$, explicitly, if one can write the potential as
\begin{equation}
	V(\phi) = V_0 + \bar{V}(\phi),
	\label{eq:genericVdecomp}
\end{equation}
we are in the regime where $\bar{V}(\phi)/V_0 \ll 1$ for $\phi\in(\phi_-,\phi_+)$. Gravitational effects can be compared to tunneling phenomena using the expansion parameter $H$ and the variation scale of the potential between the minima, $\bar{V}\sim m^4$, and considering its ratio. So that with $H^2\sim V_0/M_{\rm Pl}^2$ from the first Friedmann equation, the ratio $(H/m)^4 \sim V_0^2/(m M_{\rm Pl})^4$ can surpass one if $V_0/M^4_{\rm Pl} \gtrsim m^4/V_0$ without leaving the de Sitter regime. Similar studies have been carried out in the past, in particular, we expand on the work done by  \cite{Dunne2006}, where the starting Euclidean action is
\begin{align}
	S^{\rm Dunne} = \int \d^4 x \sqrt{-g}\left[ \frac{1}{2} \nabla_\mu\phi\nabla^\mu \phi - V(\phi) - \frac{1}{2\kappa}R \right],
\end{align}
 $R$ being the Ricci scalar and $\kappa=8\pi/M_{\rm Pl}^2$. However, we focus on the special case $\kappa\ll 1$, which corresponds to gravitational decoupling explained above and ignore any backreaction on the geometry. With this in mind, we use the simplified action:
\begin{align}
	S &= \int\dd[4]{x} \sqrt{-g} \left[\frac{1}{2}g^{\mu\nu}\partial_{\mu}\phi\partial_{\nu}\phi - V\left(\phi\right)\right],
	\label{eq:originalAction}
\end{align}
where $\phi$ is a real scalar field, $g_{\mu\nu}$ is the metric corresponding to de Sitter space. The potential is chosen to be polynomial and to feature two different local minima
\begin{equation}
V(\phi) = V_0 - \frac{m^2}{2}\phi^2 -\frac{b}{3!}\phi^3 + \frac{\lambda}{4!}\phi^4.
\label{eq:potential}
\end{equation}
We Wick rotate the entire action by performing the substitution $t \rightarrow \iu \tau$,
\begin{equation}
S[\phi] \rightarrow -\iu \int \dd\tau \dd[3]\vec{x} \sqrt{g_E}\bigg[\frac{1}{2}g^{ij}_{E}\,\partial_i \phi\partial_j\phi + V(\phi)\bigg]\equiv - S_E[\phi],
\label{eq:euclideanActionFull}
\end{equation}
denoting by $g_E$ the positive-definite Euclidean metric. This has the effect of creating a saddle point in the path integral, while at the same time changing the sign of the potential.
To understand the gradients of $\phi$, $\partial_{\mu}\phi$, we expand the action around our background
\begin{align}
\phi & \rightarrow \phi^{(0)} + \phi^{(1)},
\end{align}
where $\phi^{(0)}$ is the solution to the tree-level equation of motion, with appropriate boundary conditions, which will be discussed later in Sec.\ref{subsec:bounce}. The choice of parametrization of the fluctuations above is equivalent to choosing the weighing function for fluctuations also used in  \cite{Dunne2006}:
\begin{align}
	||\phi^{(1)}||^2 \equiv \int \d^4 x\, \sqrt{g_{\rm E}} (\phi^{(1)})^2 = 2\pi^2\int \d\tau\, a^3(\tau)\, (\phi^{(1)})^2.
\end{align}
This fixes the form of the fluctuation operator we use for subsequent computations (for alternatives, consider the appendices in  \cite{Dunne2006} and  \cite{Koehn:2015hga}).
The expansion is
\begin{align}
\begin{aligned}
S_E &
= \int\dd[4]{x} \sqrt{g_E} \bigg[\frac{1}{2}g^{ij}_E\,\partial_{i}\phi^{(0)}\partial_{j}\phi^{(0)} + V\left(\phi^{(0)}\right) + g^{ij}_E\, \partial_{i}\phi^{(0)}\partial_{j}\phi^{\left(1\right)} + \eval{\pdv{V}{\phi}}_{\phi=\phi^{(0)}}\phi^{(1)}\\
 &\qquad  + \frac{1}{2}g^{ij}_E\,\partial_{i}\phi^{\left(1\right)} \partial_{j}\phi^{(1)} + \frac{1}{2}\eval{\pdv[2]{V}{\phi}}_{\phi=\phi^{(0)}} \phi^{\left(1\right)}\phi^{(1)}\bigg].
\end{aligned}
\end{align}
with Latin indices denoting Euclidean coordinates, i.e. $i,j=1,2,3,4$ and $\tau$ understood as $x^4$. Here linear terms in $\phi$ vanish by construction. The derivatives of the potential are explicitly
\begin{align}
\frac{\partial V(\phi)}{\partial \phi} & =-m^{2}\phi - \frac{b}{2}\phi^2 +\frac{\lambda}{6}\phi^{3},\\
\frac{\partial V^2(\phi)}{\partial \phi^2} & =-m^{2} - b\phi +\frac{\lambda}{2}\phi^{2}.
\end{align}
We split the action $S_E=B+S_E^{(2)}\left[\phi^{\left(1\right)}\right]$,
where $S_E^{(2)}\left[\phi\right]$ contains bilinear terms for $\phi$, with operators evaluated over the $\phi^{(0)}$ background. Higher dimensional terms do not contribute at the one-loop level, so that
\begin{align}
B &\equiv \int \dd[4]{x}\, \sqrt{g_E}\, \left[\frac{1}{2}g^{ij}_E\,\partial_{i}\phi^{(0)}\partial_{j}\phi^{(0)} + V\left(\phi^{(0)}\right)\right]\\
S_E^{(2)}[\phi^{(1)}] &\equiv \int \dd[4]{x}\,\sqrt{g_E}\,\left[\frac{1}{2}g^{ij}_E\,\partial_{i}\phi^{(1)}\partial_{j}\phi^{(1)} + \frac{1}{2}\eval{\pdv[2]{V}{\phi}}_{\phi=\phi^{(0)}}\phi^{(1)}\phi^{(1)} \right].
\end{align}
Truncating $S_E$ to second order, we obtain a Gaussian path integral
\begin{align}Z\left[0\right] & =\int{\cal D}\phi\exp\left(-\frac{1}{\hbar}B-\frac{1}{\hbar}S_E^{(2)} \right) =
\exp\left(-\frac{1}{\hbar}B\right)\int{\cal D}\phi\exp\left(-\frac{1}{\hbar}S_E^{(2)}\right).
\label{eq:partitionFunction}
\end{align}
We can compute the path integral containing the fluctuations as follows for the case of de Sitter, using Eq.~\eqref{eq:volumeElement}:
\begin{align}
S_E^{(2)}\left[\phi^{\left(1\right)}\right] & =\int \d{\Omega_3}\d\tau  \left[-\frac{1}{2}\phi^{(1)}\partial_{i}\left(a^3(\tau) \,g^{ij}_E\,\partial_{j}\phi^{(1)}\right)+ a^3(\tau) \,\frac{V^{\prime\prime}\left(\phi\right)}{2}{\phi^{(1)}}^{2}\right]\\[7pt]
&=\int \d{\Omega_3}\d\tau \, \left[ \frac{1}{2}\phi^{(1)} \left[-\partial_{i} a^3(\tau) \,g_{E}^{ij}\,\partial_{j} + a^3(\tau)\, \frac{\partial^2 V(\phi)}{\partial \phi^2}\right]\phi^{\left(1\right)}\right],
\label{eq:actionOneLoopEffs}
\end{align}
where an integration by parts was performed to obtain the first line. We then define the fluctuation operator as
\begin{align}
{\cal G}_{b}^{-1}(x,y) &\equiv \delta(x-y)\left(-\partial_{i} a^3(\tau)\,g_{E}^{ij}\partial_{j} + a^3(\tau)\,\eval{\frac{\partial^2 V}{\partial \phi^2}}_{\phi=\phi^{(0)}}\right)
\label{eq:fluctuationsOpGeneral}
\end{align}
and include a regularization for the determinant by evaluating at the false vacuum,
\begin{align}
{\cal G}_{-}^{-1}\left(x,y\right) & = \delta\left(x-y\right) \left(-\partial_{i}a^3(\tau) \,g^{ij}_E\,\partial_{j} + a^3(\tau)\,\eval{\frac{\partial^2 V}{\partial \phi^2}}_{\phi=\phi_-}\right).
\end{align}
We thus obtain finite results at the price of introducing an additional contribution equivalent to a shift in the cosmological constant. In contrast to the flat space case, where this is actually just an unphysical constant, we assume that this results from the backreaction on the geometry and do not focus our attention on the consequences or physical interpretation of this shift. Using the result of a Gaussian path integral we obtain schematically
\begin{align}
Z\left[0\right] &\propto \left|\frac{\det'\left({\cal G}_{b}^{-1}\left(x,y\right)\right)}{\det\left({\cal G}_{-}^{-1}\left(x,y\right)\right)}\right|^{-\frac{1}{2}}\ee^{-B},
\end{align}
where the prime indicates the determinant does not include negative or zero modes. The proportionality factor will be discussed later, when we examine the non-positive modes.

In a more general setting, where gravitational effects are included, the fluctuations of the metric can lead to non-trivial constraints on the fluctuation operator. Expressions for such cases have been found  \cite{Gratton1999,Gratton:2001Hmo}, from which we can specialize to our de Sitter case. The parametrization we use follows Dunne's previous work  \cite{Dunne2005,Dunne2006}, where the following expressions are derived while neglecting fluctuations of the scale factor and fixing its evolution to be as in de Sitter space, namely where the scale factor is fixed as in Eq.~\eqref{eq:scaleFactordS}.

As it is shown in the literature  \cite{Allen:1983Pti,Hawking1983,Vilenkin1983, Birrell:1982ix, Markkanen2018T1l}, a theory with a scalar field with self-interactions over de Sitter spacetime is renormalizable, but as opposed to the flat case scenario, where there is a clear interpretation for the divergent pieces in terms of counterterms of the scalar couplings, the scalar sector one-loop contribution now demands as well the removal of purely gravitational divergent pieces.

\subsection{Computing the classical background}
\label{subsec:bounce}

In order to compute the decay rate of the false vacuum, we must first find the saddle-point field configuration known as the bounce, $\phi^{(0)}$. Motivated by the proofs of minimization in the flat space cases \cite{Coleman1977}, we look for an $O(4)$ symmetric solution to the equation of motion, which greatly simplifies to
\begin{equation}
	\ddot{\phi}+ 3 H\cot(H\tau)\dot{\phi}=V'(\phi),
	\label{eq:eomBounce}
\end{equation}
where the dot denotes differentiation with respect to $\tau$ and $\prime$ denotes derivatives with respect to $\phi$.
It is convenient to employ the dimensionless field $\frac{\phi}{v}$; however, in order to avoid complicating the notation, we keep using $\phi$, while it is to be understood to be written in units of $v$ for the rest of the document unless otherwise specified. We can now write the potential as
\begin{align}
	V\left(\phi\right) &= V_{0} + \beta H^{2}v^{2} \left(-\frac{1}{2}\phi^{2}-\frac{b}{3}\phi^{3}+\frac{1}{4}\phi^{4}\right),
	\label{eq:rescaledPot}
\end{align}
with the tree-level mass parameter, $m^2$, reparametrized as $\beta H^2$ and the equation of motion as
\begin{equation}
	\ddot{\phi}+ 3 \cot(\sigma)\dot{\phi}=\frac{V'(\phi)}{H^{2}v^{2}},
	\label{eq:eomBounceRescaled}
\end{equation}
where, in this instance, the dots denote derivatives with respect to the rescaled time $\sigma=H\tau$ and prime is a derivative with respect to the new $\phi$.

The bounce has the following boundary conditions, $\dot\phi_b(0)=0$ and $\phi_b(\pi)=\phi_-$ for $\phi$ as a function of $\sigma$. These are obtained from the Cartesian analogous conditions presented in Sec.~\ref{sec:vacuumDecayFlat} simply transcribed to our de Sitter coordinates. Notice that in the de Sitter case without a thin wall approximation, there is a friction term in the equation of motion Eq.~\eqref{eq:eomBounce}, which changes sign within $(0,\pi)$. Thus by continuity (of an under/overshooting procedure), a bounce solution with the above boundary conditions exists\footnote{A posteriori we verify that a bounce with $\phi(0)=\phi_-$, does indeed exist.}. Moreover, in contrast to the flat space case, where the bounce solution does not actually reach the true vacuum when a thick wall is considered, the present bounce configuration does. This is a special feature of this model, where the geometry is kept fixed to de Sitter, and will not hold  in a more general setting.

\section{Functional determinant of the fluctuation operator}
\label{sec:funcDets}

When we examine the fluctuation operator, specializing to the Euclidean de Sitter space, from Eq.~\eqref{eq:actionOneLoopEffs}, we get:
\begin{equation}
	S_E^{(2)}\left[\phi^{\left(1\right)}\right] = \frac{1}{2}\int \d\Omega_3 d\tau \phi^{\left(1\right)}\bigg[-\partial_\tau a^3(\tau)\partial_\tau + a^3(\tau) U(\tau)\bigg] \phi^{\left(1\right)},
	\label{eq:action2}
\end{equation}
with
\begin{align}
	U(\tau)= V''(\phi^{(0)}) - \frac{\Delta_3}{a^2},
\end{align}
where $\Delta_3$ stands for the Laplacian on the 3-sphere.
The fluctuations operator for the general case, when variations of the metric are considered, can be found in  \cite{Dunne2006, Gratton1999}, nonetheless in this document we are only interested in the case where the background is fixed to be de Sitter space, and the  fluctuations operator is then explicitly
\begin{align}
	\delta^4(x-x') \mathcal{G}^{-1}_b(x,x') \equiv \delta^4(x-x')\frac{\sin^3(H\tau)}{H^{3}}\Bigg( - \partial_\tau^2 - 3H\cot(H\tau) \partial_\tau - H^2\csc^2(H\tau)\Delta_3 + V''(\phi^{(0)})\Bigg).
	\label{eq:flucOp}
\end{align}
The related differential equation for a scalar field fluctuation written in terms of hyper-spherical harmonics (for notation and properties see \cite{Higuchi1987}) gives the Jacobi equation:
\begin{equation}
	\ddot{\phi}^{(1)} + 3\frac{\dot{a}}{a}\dot{\phi}^{(1)} - \left[V''(\phi^{(0)})
	+ \frac{\ell(\ell+2)}{a^2}\right]\phi^{(1)} = \lambda\phi^{(1)}.
	\label{eq:jacobi}
\end{equation}

\subsection{Negative and zero modes}
\label{subsec:zeromodes}
The fluctuations operator ${\cal G}_{b}^{-1}$ may have zero modes, in other words eigenfunctions for the fluctuation operator with eigenvalue zero. It is important to notice that these modes, in this case, follow a wave equation with an effective mass, nevertheless they represent directions in field space in which one may deform the scalar background configuration without modifying the value of the action up to one-loop order. Zero modes if present, have to be exchanged by collective coordinates in order to avoid the functional determinant being immediately zero. A typical zero mode is usually related to translations of the bounce $\partial_{\mu}\phi^{(0)}$. If $\phi^{(0)}$ satisfies
\begin{align}
\frac{1}{\sqrt{g_E}}\,\partial_{i}\left(\sqrt{g_E}\,g^{ij}_E\,\partial_{j}\phi^{(0)}\right)
 &= V^{\prime}\left(\phi^{(0)}\right),
	\label{eq:bounceEqGeneral}
\end{align}
it is usually possible to show that translations along the directions of the wall of the bounce satisfy the fluctuation operator differential equation with a zero-eigenvalue. This is done by computing and additional derivative of the equation of motion, viz.
\begin{align}
0 & =\partial_{k}\left\{ \frac{1}{\sqrt{g_E}\,}\partial_{i}\left(\sqrt{g_E}\,g^{ij}_E\,\partial_{j}\phi^{(0)}\right) - V^{\prime}\left(\phi^{(0)}\right)\right\} \\[6pt]
 &= -\left(\frac{1}{2}g_E^{-3/2}\, \partial_{k} g_E\right)\partial_{i} \left(\sqrt{g_E}\,g^{ij}_E\right) \partial_{j}\phi^{(0)}
 - \left(\frac{1}{2 g_E}\, \partial_{k} g_E\right) g^{ij}_E \partial_{i}\partial_{j}\phi^{(0)}\notag\\
 &\qquad + \frac{1}{\sqrt{g_E}\,} \partial_{i}\left[\partial_{k}\left(\sqrt{g_E}\,g^{ij}_E\,\right)\partial_{j}\phi^{(0)}\right] + \frac{1}{\sqrt{g_E}\,} \partial_{i}\left(\sqrt{g_E}\,g^{ij}_E\, \partial_{j}\partial_{k}\phi^{(0)}\right) -V''\left(\phi^{(0)}\right)\partial_k \phi^{(0)}.
\end{align}
Since the metric has some dependencies on the angular coordinates, the first three terms are non-zero, and we cannot immediately conclude that partial derivatives of the bounce configuration become zero modes. In the zero temperature flat spacetime case, the $SO(4)$-symmetric bounce is expressed through the coordinate $r^2=\tau^2+x^2+y^2+z^2$ so that such a configuration still depends on each direction indirectly and $\partial_i\phi^{(0)}$ is not immediately null for $i\in\{\tau,x,y,z\}$. In our case, the bounce is expressed in terms of the independent coordinate $\tau$ and translations with respect to any other direction are by construction zero. This leads to an interesting question: what happens to the zero modes in this scenario?

Given that we are not working under the thin wall approximation and the temporal direction always remains compact once the metric is Euclideanized, there is the possibility that the zero modes are no longer present, because fixing the background in this manner breaks the symmetry. We can then consider expanding an arbitrary fluctuation in a basis of hyper-spherical harmonics as the fluctuation operator can still be solved for through a separation of variables. Then one can check every $\ell$ sector for zero modes in the radial direction. This we do for our benchmark point in the numerical implementation described in Sec.~\ref{sec:numericalImpl}. In our numerical treatment, we consider the sector $\ell=1$ and give some of our insights and results on this point which was left unanswered in  \cite{Dunne2006}.

The lack of zero modes seems to impact the interpretation of the mass dimensions of the expressions. In its most naive saddle-point expansion, the decay rate acquires its dimensions via the extraction of the zero modes in the fluctuation determinant evaluated at the bounce (read pertaining comments in\cite{Callan1977}). If this is not the case, full cancellation of eigenvalues occurs when taking the ratio with the functional determinant of the fluctuation operator evaluated at the false vacuum. From this perspective, we may only interpret the imaginary part of the action as a tunneling probability, instead of a tunneling probability per spacetime volume. We do not include the contributions of the sector $\ell=1$ in the following computations, as we do not have a general proof of the absence or presence of zero modes.

Besides the zero modes, related to the translation symmetry of the problem, we know that there must be at least one mode of negative eigenvalue present in order to give the appropriate imaginary contribution to the energy in the tunneling process. It was shown by Coleman that for the present case there is only a single negative mode, which lies in the $\ell=0$ sector \cite{Coleman88}. In order to extract it we therefore again consider the fluctuation equation setting $\ell=0$
\begin{equation}
	\ddot{\phi}^{(1)}+3\frac{\dot{a}}{a}\dot{\phi}^{(1)} - V''(\phi^{(0)})\phi^{(1)}=\lambda\phi^{(1)}.
	\label{eq:negmode}
\end{equation}
Since we have compactified the temporal direction, we expect the spectrum of the fluctuation operator to be discrete, and since there is only one negative mode, it will belong to the fluctuation with the lowest eigenvalue. We find the lowest eigenvalue numerically for the benchmark case illustrated in Sec.~\ref{sec:numericalImpl}.

\subsection{The determinant of the fluctuation operator via the Gel'fand Yaglom theorem}

The eigenvalue equation of the fluctuations operator $G_b^{-1}$ as defined through Eq.~\eqref{eq:flucOp} cannot be solved directly. Instead we make use of the  Gel'fand-Yaglom \cite{Gelfand:1959nq} method to calculate the determinant ratio. After writing down the eigenfunctions as linear combinations of hyperspherical harmonics we can define functional determinants for each fixed value of $\ell$ as
\begin{equation}
	\frac{\det'\mathcal{G}^{-1}_{b,\ell}(\tau,\tau')}{\det
	\mathcal{G}^{-1}_{-,\ell}(\tau,\tau')} =
	\frac{\Phi^{(\ell)}(\tau_{\text{f}})}{\Phi^{(\ell)}_{-}(\tau_{\text{f}})},
	\label{eq:GelfandYaglom}
\end{equation}
where $\Phi^{(\ell)}$ are the eigenfunctions for the eigenvalue $0$ for their corresponding fluctuation operator. Their calculation is greatly facilitated by the fact that there exists an analytical solution for the homogeneous configuration case:
\begin{equation}
	\Phi^{(\ell)}_{-}(\tau)=\frac{N}{\sin(H\tau)}P_{\alpha}^{\ell+1}\left(\cos(H\tau)\right)\qquad\text{with}\qquad \alpha = -\frac{1}{2} + \iu\sqrt{\frac{V''(\phi_{-})}{H^2}-\frac{9}{4}},
	\label{eq:false-vacuum-fluctuation}
\end{equation}
where $y=P^{\ell +1}_\alpha(x)$ are the associated Legendre functions of the first kind and solve
\begin{align}
	\left(1 - x^2\right) y'' - 2xy' + \left[\ell(\ell+1) - \frac{\alpha^2}{1-x^2}\right] y = 0.
\end{align}
With the definition of the auxiliary function
\begin{equation}
	T^{(\ell)}(\tau) :=\frac{\Phi^{(\ell)}(\tau)}{\Phi^{(\ell)}_{-}(\tau)},
	\label{eq:definitionT}
\end{equation}
it is necessary to solve the differential equation for each $\ell$ sector
\begin{equation}
	-\ddot{T}^{(\ell)}(\tau)-\left[2\frac{\dot{\Phi}^{(\ell)}_{-}}{\Phi^{(\ell)}_{-}} + 3\frac{\dot{a}}{a}\right]\dot{T}^{(\ell)}(\tau) + \left[ U(\phi^{(0)}) -\bigg(V''(\phi_-)-\frac{\Delta_3}{a^2}\bigg) \right] T^{(\ell)}(\tau) = 0,
	\label{eq:ratio-operator}
\end{equation}
subject to initial conditions $T(0)=1$ and $\dot{T}(0)=0$, to obtain the ratio of functional determinants:
\begin{equation}
	T^{(\ell)}(\tau_\text{f}) =	\frac{\det'\mathcal{G}^{-1}_{b,\ell}(\tau,\tau')}{\det\mathcal{G}^{-1}_{-,\ell}(\tau,\tau')}.
\end{equation}
Knowing the solutions to Eq.~\eqref{eq:ratio-operator} for all values of $\ell$ enables us to compute the functional determinant in theory through the following expression for the one-loop corrections:
\begin{equation}
	\ln \frac{\det' \mathcal{G}^{-1}_{b}(x,y)}{\det
	\mathcal{G}^{-1}_{-}(x,y)} = 	\tr' \log\frac{
	\mathcal{G}^{-1}_{b}(x,y)}{\mathcal{G}^{-1}_{-}(x,y)} = \sum_{\ell=2}^\infty (\ell + 1)^2\ln T^{(\ell)}(\tau_f),
	\label{eq:totaldet}
\end{equation}
where the prime, next to the determinant and the trace, indicates that negative and  possible zero modes are to be omitted. In order to get the last equality, the angular part has been factorized and written explicitly. In the practice we find the $T^{(\ell)}$ functions numerically for some values of $\ell$, more details are given later in section \ref{sec:numericalImpl}.

\subsection{The determinant of the fluctuation operator via Green's functions}
We can define a Green's equation by employing the fluctuations operator from Eq.~\eqref{eq:flucOp}:
\begin{align}
	\mathcal{G}^{-1}_b(x)\mathcal{G}(x,x') &= \delta(x-x').
\end{align}
The usual spectral decomposition can be written as:
\begin{align}
	\mathcal{G}(x,x') = \sum_\lambda \frac{1}{\lambda} f_\lambda(x)\, f^*_\lambda(x'),
	\label{eq:spectralDecompG}
\end{align}
where the $\lambda$'s are the eigenvalues of the operator $\mathcal{G}^{-1}$ that may be discrete, continuous or mixed and $f_\lambda(x)$ are the corresponding eigenfunctions. The Euclidean de Sitter space is compact and immediately implies that the spectrum will be formally discrete. Given the symmetry of the problem, it is possible to further factorize each eigenfunction into a $\tau$-dependent piece and an angular part. By taking the latter to be a hyperspherical harmonic $Y_{\ell,m_1,m_2}$, we can diagonalize the Laplace-Beltrami operator on $S^3$,
\begin{align}
	-\Delta_3 Y_{\ell,m_1,m_2}(\theta_1,\theta_2,\varphi) =  \ell(\ell+2)Y_{\ell,m_1,m_2}, \qquad \text{with}\qquad \ell\geq m_1\geq |m_2|.
\end{align}
By taking the eigenfunctions to be  $f_\lambda(x) = \psi_n(\tau)Y_{\ell,m_1,m_2}(\theta_1, \theta_2, \varphi)$, the Green's function can be written as
\begin{align}
	\mathcal{G}(x,x') = \sum_{n,\ell\geq m_1\geq |m_2|}\frac{1}{\lambda_{n,\ell}} \psi_n(\tau)\psi_n^*(\tau') Y_{\ell,m_1,m_2}(\theta_1,\theta_2,\varphi)Y^*_{\ell,m_1,m_2}(\theta'_1,\theta'_2,\varphi').
\end{align}
The sums over over $m_1$ and $m_2$ can be carried out using the sum rules for hyper-spherical harmonics \cite{Avery1982}, to obtain
\begin{align}
\sum_{m_1\geq |m_2|}
Y_{\ell,m_1,m_2}(\theta_1,\theta_2,\varphi)Y^*_{\ell,m_1,m_2}(\theta'_1,\theta'_2,\varphi')
 = \frac{(1+\ell)}{2\pi^2} C_\ell^{1}({\bf e}\cdot{\bf e'}) =
\frac{(1+\ell)}{2\pi^2} U_\ell({\bf e}\cdot{\bf e'}).
\label{eq:angularPartGreens}
\end{align}
Here ${\bf e}$ and ${\bf e}'$ are unit vectors associated with the angle coordinates $(\theta_1,\theta_2,\varphi)$ and $(\theta'_1,\theta'_2,\varphi')$ respectively and $C_n^{(m)}(x)$ are Gegenbauer polynomials which coincide with Chebyshev polynomials of the second kind, $U_n$,  for $m=1$. This leads to the following simplification of the Green's function
\begin{align}
	\mathcal{G}(x,x') = \frac{1}{2\pi^2}\sum_{n,\ell}\frac{(1+\ell)}{\lambda_{n,\ell}}
	U_\ell({\bf e}\cdot{\bf e'})\psi_n(\tau)\psi^*_n(\tau') \equiv
	\frac{1}{2\pi^2}\sum_{\ell}(1+\ell) U_\ell({\bf e}\cdot{\bf e'})
	\mathcal{G}_{\ell}(\tau,\tau').
	\label{eq:fullGreenSolution}
\end{align}
With this factorization, we obtain the following Green's function equation for the $\tau$ part from Eq.~\eqref{eq:flucOp}:
\begin{align}
\frac{\sin^3(H\tau)}{H^3}\left( - \partial_\tau^2 - 3H\cot(H\tau) \partial_\tau  + H^2\csc^2(H\tau)\ell(\ell+2)  + V''(\phi^{(0)})\right)\mathcal{G}_{\ell}(\tau,\tau')
&= \delta(\tau-\tau').
\end{align}
We can simplify the computation if we consider the \emph{alternative Green's function equation}
\begin{align}
\left(- \partial_\tau^2 - 3H\cot(H\tau) \partial_\tau  + H^2\csc^2(H\tau)\ell(\ell+2)  + V''(\phi^{(0)})\right)G_{\ell}(\tau,\tau')
	&= \delta(\tau-\tau'),
	\label{eq:reducedFluctuationOp}
\end{align}
and later correct for the factor of $\sin^3(H\tau)/H^3$ multiplying the Dirac delta as we explain in the following.

The alternative equation is numerically more stable and we are able to compute the corresponding Green's functions up to values of $\ell=200$, an example of the results is shown in Fig.\ref{fig:sampleCoinGreen} in the numerical implementation section~\ref{sec:numericalImpl}. The Green's functions are computed using an Ansatz which decomposes the two-point function $G_{\ell}(\tau,\tau')$ into an increasing and a decreasing solution to the left and to the right of $\tau'$, explicitly
\begin{align}
	G_{\ell}(\tau,\tau') =
	\frac{1}{\mathcal{W}[g_R,g_L]}\bigg[\Theta(\tau-\tau')g_{\ell,\,
	L}(\tau)g_{\ell,\, R}(\tau') +
	\Theta(\tau'-\tau)g_{\ell,\, L}(\tau')g_{\ell,\, R}(\tau)\bigg],
	\label{eq:ansatzReducedGreenFunctions}
\end{align}
where $\Theta$ is the Heaviside step function and $\mathcal{W}$ is the Wronskian,
\begin{align}
	\mathcal{W}[f,g](x) = \det\begin{pmatrix}
		f(x) & g(x)\\
		f'(x) & g'(x)
	\end{pmatrix}.
\end{align}
This decomposition already considers continuity and the appropriate jump in the first derivative, reducing the problem to two simpler one-dimensional problems.
That is, $g_{L,R}$ must solve the homogeneous version of the differential Eq.\ref{eq:reducedFluctuationOp} with boundary conditions $g_L(0)=0, g_L'=1, g_R(\pi/H)=0$ and $g_R'(\pi/H)=-1$ for every value of $\ell$.
It is worth noting that the value taken for the first derivative is auxiliary, and the result does not depend on it, as expected for a second order differential equation.
We can recover the actual Green's functions correcting the Wronskian of the alternative equation by reintroducing the scale factor, $a^3(\tau)=\sin^3(H\tau)/H^3 $, that was ignored, explicitly
\begin{align}
	\mathcal{G}_{\ell}(\tau,\tau') =
	\frac{1}{a^3(\tau')\mathcal{W}[g_R,g_L]}\bigg[\Theta(\tau-\tau')g_{\ell,\,
		L}(\tau)g_{\ell,\, R}(\tau') +
	\Theta(\tau'-\tau)g_{\ell,\, L}(\tau')g_{\ell,\, R}(\tau)\bigg].
	\label{eq:ansatzGreenFunctions}
\end{align}

Once the value of $\ell$ is specified and the bounce solution is available, it is possible to use available integration routines to obtain $G_\ell(\tau,\tau')$ numerically.
We expand and comment on the numerical implementation in Sec.~\ref{sec:numericalImpl}.
We only need to connect the Green's functions with the functional determinant.
This can be done by means of the resolvent method \cite{Baacke:1993aj, Baacke:1993jr}.
In brief, for a positive-definite operator, it is possible to write
\begin{align*}
	\log\frac{\mathcal{G}^{-1}_b(x,y)}{\mathcal{G}^{-1}_-(x,y)} &= \int \d\lambda\,\log(\lambda)f_{\lambda,b}^{\phantom{*}}(x) f_{\lambda,b}^*(y)  - \int \d\lambda\,\log(\lambda) f_{\lambda,-}^{\phantom{*}}(x) f_{\lambda,-}^*(y) \\[6pt]
	&=  - \int\d\lambda \left(\int_0^\infty \d s\, \frac{g_{\lambda,b}^{\phantom{*}}(x) g_{\lambda,b}^*(y)}{\lambda + s} - \int_0^\infty \d s\, \frac{f_{\lambda,-}^{\phantom{*}}(x) f_{\lambda,-}^*(y)}{\lambda + s}\right)\\[6pt]
	&= -\int_0^\infty \d s\, \left( \mathcal{G}_{b_s}(x,y) - \mathcal{G}_{-_s}(x,y) \right),
\end{align*}
where
\begin{equation}
	\mathcal{G}_s(x,y) \equiv  \int \d\lambda \frac{f_\lambda(x)f_\lambda^*(y)}{\lambda + s}
	\label{eq:greenSpectralDecomp}
\end{equation}
are the Green's functions associated with the deformed operator
\begin{align}
	\mathcal{G}_s^{-1}(\tau,\tau') \equiv \mathcal{G}^{-1}(\tau,\tau') + s\mathds{1}
	\label{eq:deformedOp}
\end{align}
and where the integral over $\lambda$ above may refer to a sum in the discrete case, we however will approximate the result by assuming a continuous spectrum in the numerical implementation. For our specific case we arrive at
\begin{align}
	\log\frac{\det' \mathcal{G}^{-1}_b(x,y)}{\det\mathcal{G}^{-1}_-(x,y)} &=
	\tr'\log\frac{\mathcal{G}^{-1}_b(x,y)}{\mathcal{G}^{-1}_-(x,y)} = \tr'\log\frac{a^3(\tau)G^{-1}_b(x,y)}{a^3(\tau)G^{-1}_-(x,y)}.
\end{align}
For positive-definite operators the combination $\tr\log$ satisfies the usual properties of the logarithm and linearity, so that the scale factors $a$, cancel and we can employ the following formula to obtain the determinant using simply the alternative Green's functions
\begin{align}
	\frac{1}{2}\log\frac{\det' \mathcal{G}^{-1}_b(x,y)}{\det\mathcal{G}^{-1}_-(x,y)}
	&= -	\frac{1}{2}\sum_{\ell=2}^\infty (\ell+1)^2\int_0^{\pi/H}\d \tau\int_0^\infty \d s\, \left( G_{b_{(\ell,s)}}(\tau,\tau) - G_{-_{(\ell,s)}}(\tau,\tau) \right),
	\label{eq:logDetToGreens}
\end{align}
where the negative and possible zero modes have been omitted.
Usually, in the flat-space case and the thin-wall approximation (see Refs.~ \cite{Ai:2018guc, Ai:2020sru}), the knowledge of the Green's function for all values of the momentum along the bubble wall means that the Green's functions for the deformed operator are readily available. Here in the de Sitter case, because of the $\csc^2(H\tau)$ factor in the fluctuation operator, that is no longer the case and an additional scan over the $s$ parameter is required to obtain the one-loop term.

\section{Renormalization}
\label{sec:renorm}

It is known that UV divergences in de Sitter spacetime can be dealt with by means of dimensional regularization \cite{Birrell:1982ix} in the more general case where the metric is not fixed.
As we are considering a scenario where the de Sitter background is fixed, we renormalize instead by following the techniques described by Dunne et al.  \cite{Dunne2005} by means of the WKB approximation and extend the method to define a homogeneous fraction of the one-loop contributions, which will allow us to determine the gradient contributions to the one-loop level. In the following section, similar methods are employed to obtain a renormalized tadpole function and a quantum corrected bounce.

\subsection{Applying the WKB Ansatz to the Jacobi equation}
Let us consider the Jacobi equation, Eq.~\eqref{eq:jacobi}.
To extract the divergent behavior we consider its large-$\ell$ limit.
First, we need to bring this expression into a form such that the WKB method can be directly applied.
To do so, we absorb the linear term in the derivatives to obtain a Schr\"odinger-type equation.
Consider therefore an equation of the type
\begin{equation}
	\ddot{\varphi}(\tau) + \omega^2(\tau)\varphi(\tau) = 0,
\end{equation}
and define
\begin{equation}
	\varphi(\tau) \equiv c(\tau)\phi^{(1)}(\tau).
\end{equation}
As we want to solve for $\omega$, we insert this into the expression above to obtain
\begin{equation}
	c(\tau)\ddot{\phi}^{(1)}(\tau) + 2\dot{\phi}^{(1)}(\tau)\dot{c}(\tau) + \ddot{c}(\tau)\phi^{(1)}(\tau) + \omega^2c(\tau)\phi^{(1)}(\tau) = 0.
\end{equation}
We identify
\begin{align}
	2\frac{\dot{c}}{c} &= 3\frac{\dot{a}}{a},\quad\text{and}\quad
	\frac{\ddot{c}}{c}+\omega^2 = -U,
\end{align}
then comparing coefficients, straightforwardly yields the frequency:
\begin{equation}
\omega^2=-\left(\frac 34 \frac{\dot{a}^2}{a^2}+\frac 32 \frac{\ddot{a}}{a}+U(a,\phi,\ell)\right).
\end{equation}
The Jacobi equation for the false vacuum solution is almost the same, with the exception of a different fluctuation potential
\begin{equation}
U_{-}\equiv V''(\phi_{-})+\frac{\ell(\ell+2)}{a^2}.
\end{equation}
Thus,
 \begin{equation}
\omega_{-}^2 = -\left(\frac 34 \frac{\dot{a}^2}{a^2}+\frac 32 \frac{\ddot{a}}{a}+U_{-}\right).
\label{eq:freq}
\end{equation}
For the fluctuation alone, the Ansatz to solve equation\eqref{eq:jacobi} is given by
\begin{equation}
	\phi^{(1)}=\frac{1}{\sqrt{\omega}}\ee^{\pm i \int_0^{\tau_{\text{max}}} d \tau \:\omega},
	\label{eq:wkbAnsatz1}
\end{equation}
where $\omega$ takes the form stated earlier. The validity of the WKB approximation is subject to the condition that
\begin{equation}
	\abs{\frac{\dot{\omega}}{\omega^2}}\ll 1,
\end{equation}
which is satisfied for all large values of $\ell$.
We expect the accuracy of the Ansatz to increase with growing $\ell$, as this corresponds to a larger effective mass for the scalar field. Using now the definition of $T$ in Eq.~\eqref{eq:definitionT}, we can write the WKB approximation as
\begin{equation}
T^{(\ell)}(\tau_{\text{max}})=\sqrt{\frac{\omega_{-}}{\omega}}\ee^{\pm i \int_0^{\tau_{\text{max}}} d \tau (\omega-\omega_{-})}.
\end{equation}
The expansion of the one-loop contributions can be expanded in negative powers of $\ell+1$ as
\begin{equation}
	\ln \frac{\det \mathcal{G}^{-1}_{b,\ell}(\tau,\tau')}{\det \mathcal{G}^{-1}_{-,\ell}(\tau,\tau')} = \sum_{\ell=0}^{\infty} (\ell+1)^2 \bigg[ \frac{\alpha}{\ell+1}+\frac{\gamma}{(\ell+1)^3}+\mathcal{O}\big((\ell+1)^{-5}\big)\bigg],
	\label{eq:logDetExpansionInEll}
\end{equation}
and then the divergent terms can be read off as explained in the following.
From the first order solution we may now extract the leading divergence and confirm the result found in  \cite{Dunne2006}:
\begin{equation}
	\alpha=\frac{1}{2}\int \d\tau a(\tau)\, \left[ V''(\phi^{(0)})-V''(\phi_{-}) \right].
	\label{eq:wkbAlpha}
\end{equation}
We can compute the frequency to next order in the WKB expansion, whose contribution to the $\log\det$ will make it accurate up to $\mathcal{O}(\ell^{-3})$, and obtain
\begin{equation}
	\omega_{(2)}=-\left( \frac 14 \frac{\ddot{\omega}_{(1)}}{\omega^2_{(1)}}-\frac 38 \frac{\dot{\omega}_{(1)}}{\omega_{(1)}^3}\right).
\end{equation}
The $\gamma$ coefficient appearing in Eq.~\eqref{eq:logDetExpansionInEll} gives
\begin{equation}
	\gamma = -\frac 18\int \d\tau a(\tau)\, \left[( V''(\phi^{(0)})-V''(\phi_{-}))(-2-2\dot{a}^2+a^2( V''(\phi^{(0)})+V''(\phi_{-}))) \right].
	\label{eq:wkbGamma}
\end{equation}

This is very similar in appearance to the factor that was postulated by Dunne on the basis of the flat space results. However, it differs by factors that would be absent in the flat space analog and is therefore numerically different.
In Sec.~\ref{sec:numericalImpl}, we further comment on the numerical implementation of this strategy.

\subsection{Applying the WKB Ansatz to build the Green's functions}

The WKB method can also be used to compute a homogeneous version of the Green's functions in a very similar way as above, but gaining some more physical insight.
Applying the WKB method to build the Green's functions in Eq.~\eqref{eq:ansatzGreenFunctions} as in  \cite{Ai:2018guc}, it is possible to find analytical expressions for the divergent pieces of the effective potential, as a function of $\tau$ and renormalize the theory as we describe in the following.
Notice that since we are looking for expressions in terms of the coordinate $\tau$, we may not use the Green's functions of the alternative fluctuation equation.

Exploiting the Ansatz in Eq.~\eqref{eq:wkbAnsatz1}, we define
\begin{align}
	g^{\rm hom}_{\ell,\,\rm L/R} = \frac{1}{W_\ell(\tau)} \ee^{\pm
	\int_0^\tau\d\tau'\, W_\ell(\tau')}.
\end{align}
Each of the functions above must solve the homogeneous version of Eq.~\eqref{eq:reducedFluctuationOp} with boundary conditions set at opposite ends of
the range of $\tau$. The decomposition made in Eq.~\eqref{eq:ansatzGreenFunctions} already ensures that the Green's function contains the correct discontinuity in the first derivative. However, in order to find the divergent terms, we may ignore the linear derivative and assume that $\ell\gg 1$. Plugging the expression above into the deformed homogeneous differential equation (including the $s$ parameter), leads to
\begin{align}
	0 &= \ell (\ell+2) \csc^2(H \tau)+V''(\phi^{(0)}(\tau)) + s - W_\ell (\tau)^2 +
	\frac{\ddot{W}_\ell(\tau)}{2 W_\ell
		(\tau)}-\frac{3 \dot{W}_\ell (\tau)^2}{4 W_\ell (\tau)^2},
\end{align}
 where the prime denotes derivatives with respect to $\phi$ and the dots stand for
 differentiation with respect to $\tau$. Looking for a homogeneous solution, we discard the last two terms of the last equation, to find a leading order expression for
 $W_\ell$:
 \begin{align}
 	W_\ell(\tau) &= \sqrt{ \ell (\ell+2) \csc^2(H\tau)+V''(\phi^{(0)}(\tau)) + s }.
 \end{align}
Whether a closed expression for the two-point Green's function is obtainable will depend on the potential and the bounce.
For a numerical bounce, however, we can only give an expression for the coincident limit, because the exponentials in the WKB Ansätze will cancel in this limit. Eq.~\eqref{eq:ansatzGreenFunctions} reduces to
\begin{align}
	G_{b_{(\ell,s)}}^{\rm hom}(\tau,\tau) \equiv \frac{g^{\rm hom}_{\ell,\,\rm
			R}(\tau) g^{\rm hom}_{\ell,\,\rm L}(\tau)}{\mathcal{W}[g^{\rm
			hom}_{\ell,\,\rm
	R},g^{\rm hom}_{\ell,\,\rm L}]} = \frac{(\ell+1)^2}{2 \sqrt{\ell (\ell + 2)
	\csc^2(H\tau) + V''(\phi^{(0)}(\tau))+s}},
	\label{eq:homCoincidentGF}
\end{align}
where the numerator comes from taking the coincident limit of the angular part
following Eq.~\eqref{eq:angularPartGreens} and the factor of two in the denominator
comes from the Wronskian. The same expression can be found for the Green's function
evaluated at false vacuum, $\phi_{-}$, that is:
\begin{align}
	G_{-_{(\ell,s)}}^{\rm hom}(\tau,\tau) \equiv \frac{g^{\rm hom}_{\ell,\,\rm R}(\tau) g^{\rm hom}_{\ell,\,\rm L}(\tau)}{\mathcal{W}[g^{\rm hom}_{\ell,\,\rm R},g^{\rm hom}_{\ell,\,\rm L}]} = \frac{(\ell+1)^2}{2 \sqrt{\ell (\ell + 2) \csc^2(H\tau) + V''(\phi_-)+s}}.
\end{align}
We can now write a homogeneous version of the one-loop term in our curved setting, in close analogy to the well-known Coleman-Weinberg potential. Using the general relation in Eq.~\eqref{eq:logDetToGreens} we have
\begin{align}
	\begin{aligned}
	\frac{1}{2}\log\frac{\det' \mathcal{G}^{\rm hom}_{-1}(x,y)}{\det \mathcal{G}^{\rm hom}_{-1}(x,y)} &= -\frac{1}{2}\sum_{\ell=2}^\infty (\ell+1)^2\int_0^{\pi/H}\d\tau \bigg[ a^{-3}(\tau)\sqrt{\ell (\ell+2) \csc^2(H\tau)+V''(\phi^{(0)}(\tau))}\\
	&\quad - a^{-3}(\tau)\sqrt{\ell (\ell+2) \csc^2(H\tau)+V''(\phi_-)(\tau)}\bigg].
	\end{aligned}
\end{align}
Observe, that it is necessary to use the Green's function of the fluctuation operator in Eq.~\eqref{eq:fluctuationsOpGeneral} and not the ones of the alternative operator, since equality between these expressions only holds after the full trace has been performed. Since we are seeking the counterterms for the model and thus terms for the Lagrangian density, we must use the former ones. Expanding the expression for $\ell\gg 1$, or more specifically for $\ell>\sqrt{V''(\phi^{(0)})}\sim\beta$, keeping contributions up to $\ell^{-3}$ we have
\begin{align}
	\begin{aligned}
	 \frac{1}{2}\log\frac{\det' \mathcal{G}^{\rm hom}_{-1}(x,y)}{\det \mathcal{G}^{\rm hom}_{-1}(x,y)} &= \int_0^{\pi/H}\d\tau \sum_{\ell=2}^\infty \bigg[ \frac{\sin^5(H\tau)}{H^5}\frac{ V''(\phi^{(0)})^3 - V''(\phi_-)^3 }{32(\ell+1)^3}\\
	 &\quad - \frac{\sin^3(H\tau)}{H^3}\frac{3\left(V''(\phi^{(0)})^2 - V''(\phi_-)^2\right)}{32(\ell+1)^3} + \frac{\sin(H\tau)}{H}\frac{3(V''(\phi^{(0)}) - V''(\phi_-))}{32(\ell+1)^3}\\
	 &\quad  - \frac{\sin(H\tau)}{H}\frac{(V''(\phi^{(0)}) - V''(\phi_-))(\sin^2(H\tau)(V''(\phi^{(0)}) + V''(\phi_-))/H^2 - 2)}{16(\ell+1)}\\
	 &\quad + \frac{1}{4H} (\ell+1) \sin(H\tau)(V''(\phi^{(0)})-V''(\phi_-))\bigg].
	\end{aligned}
\end{align}
We can perform the sum up to a hard cutoff for $\ell$, $\ell_{\rm max}\geq \beta$, allowing us to write
\begin{align}
	\begin{aligned}
	\frac{1}{2}\log &\frac{\det' \mathcal{G}^{\rm hom}_{-1}(x,y)}{\det \mathcal{G}^{\rm hom}_{-1}(x,y)} = \int_0^{\pi/H}\d\tau\; \frac{-\psi^{(2)}(1)}{64H^5}\sin^5(\tau) \left(V''(\phi^{(0)}(\tau))^3 - V''(\phi_-)^3\right) \\
	&+\frac{1}{8H}\sin(H\tau) \bigg(\ell_{\rm max}^2[V''(\phi^{(0)}(\tau)) - V''(\phi_-)] + 3\ell_{\rm max}[V''(\phi^{(0)}(\tau)) - V''(\phi_-)]\\
	& + \log(\ell_{\rm max})[V''(\phi^{(0)}(\tau)) - V''(\phi_-)] + \left(2 + \gamma_{\rm E}  - \frac{3}{8} \psi ^{(2)}(1)\right)[V''(\phi_-) - V''(\phi_-)]\bigg)\\
	& - \frac{\sin^3(H\tau)}{16H^3} \bigg(\log(\ell_{\rm max})[ V''(\phi^{(0)}(\tau))^2 - V''(\phi_-)^2] +\big[\gamma_{\rm E} - \frac{3}{4}\psi ^{(2)}(1) \big][V''(\phi^{(0)}(\tau))^2 - V''(\phi_-)^2]\bigg).
	\end{aligned}
	\label{eq:regOneLoopContrib}
\end{align}
Here $\psi^{(m)}(z)$ is the Polygamma function defined as
\begin{align}
	\psi^{(m)}(z) \equiv \frac{\mathrm{d}^{m+1}}{\mathrm{d}z^{m+1}} \ln\Gamma(z)  =  \frac{\mathrm{d}^{m+1}}{\mathrm{d}z^{m+1}} \ln  \int_0^\infty x^{z-1} e^{-x}\,\d x.
\end{align}
The last three lines of Eq.~\eqref{eq:regOneLoopContrib} contain divergent pieces, which are to be removed by introducing counterterms to the original Lagrangian. By extracting a factor of $\sqrt{-g}$, what remains is naively expected to be a polynomial in the scalar field (if the tree-level potential was also chosen as a polynomial) so that one can then interpret the divergences to reflect the running of the self-interaction couplings, including the mass. We see from the above expression that this is not immediately the case, and although some divergent pieces have indeed the form of $\sqrt{-g}$ times a polynomial, some do not and require purely gravitational interactions to be added to the Lagrangian. As mentioned earlier, this is known from the full renormalization program in de Sitter space \cite{Birrell:1982ix, Parker:2009uva}. We perform the regularization of the expressions, nonetheless the interpretation of such divergences as local counterterms falls outside the scope of this paper.

\section{Higher order corrections from gradients}
\label{sec:higherOrder}

Gradient effects may be included in the decay rate not only through an exact numerical computation of the functional determinant over the background, but also through an additional functional derivative of the effective action, which provides a quantum corrected equation of motion for the bounce \cite{Garbrecht2015a}. It could in principle be used to reevaluate the one-loop contributions on a quantum corrected bounce, effectively including specific two-loop effects. Here we do not carry out the full program but only calculate the corrected bounce as a cross-check in relation to previous studies and a proof of concept for the de Sitter case.

Considering the effective action, shown in Eq.~\eqref{eq:effectiveAction}, but evaluated on a generic field $\phi$, we obtain the quantum equation of motion by means of a functional derivative. Let us for that purpose first rewrite the effective action more explicitly evaluated at a generic field $\varphi$
\begin{align}
	\Gamma^{(1)}[\varphi] &= S[\varphi] + \frac{\hbar}{2} \log \frac{\det \mathcal{G}^{-1}(\varphi)}{\det \mathcal{G}^{-1}(\phi_-)} = \int \d^4 x\, \sqrt{g_E}\left[ \frac{1}{2}g^{ij}_E\,\partial_{i}\varphi\partial_{j}\varphi + V_{\rm eff}(\varphi)\right]
\end{align}
where we have defined the effective potential as the sum of the tree-level potential and the one-loop contributions
\begin{align}
	V_{\rm eff}(\varphi) = V(\varphi) + V_{\rm 1-loop}(\varphi),
\end{align}
such that integrating the last term together with the spacetime volume element gives
\begin{align}
	\int \d^4 x \sqrt{g_E}\, V_{\rm 1-loop}(\varphi) = \frac{\hbar}{2} \log \frac{\det \mathcal{G}^{-1}(\varphi)}{\det \mathcal{G}^{-1}(\phi_-)}.
\end{align}
The condition above leads to the expression
\begin{align}
	\frac{\hbar}{2} \tr\log
	\frac{\mathcal{G}^{-1}(\varphi)}{\mathcal{G}^{-1}(\phi_-)} =\frac{\hbar}{2}
	\int\d\Omega_3\int \d\tau \log
	\frac{\mathcal{G}^{-1}(\varphi)}{\mathcal{G}^{-1}(\phi_-)} = \int \d^4 x
	\sqrt{g_E}\, V_{\rm 1-loop}(\varphi),
\end{align}
which allows us to identify the effective one-loop potential term as
\begin{align}
	V_{\rm 1-loop}(\varphi) = \frac{\hbar}{2}\frac{1}{a^3(\tau)}\log \frac{\mathcal{G}^{-1}(\varphi)}{\mathcal{G}^{-1}(\phi_-)}.
	\label{eq:effPotOneLoop}
\end{align}
As we have shown above, the one-loop contributions can be included in order to obtain a quantum corrected bounce:
\begin{align}
	0 &= \frac{\delta \Gamma^{(1)}}{\delta\varphi(x)} =
	\frac{\delta}{\delta\varphi(x)} \int \d^4 y\, \sqrt{g_E}\left[
	\frac{1}{2}g^{ij}_E\,\partial_{i}\varphi\partial_{j}\varphi + V_{\rm
	eff}(\varphi)\right],\\[6pt]
	0 &=
	-\frac{1}{\sqrt{g_E}}\partial_{i}\sqrt{g_{E}}\,g_{E}^{ij}\partial_{j}\varphi(x)
	+ V'(\varphi(x)) + \frac{\delta V_{\rm 1-loop}(\varphi)}{\delta
	\varphi(x)},
\end{align}
where the last term corresponds to the tadpole function for the scalar field fluctuations. This can be seen from the explicit computation of the derivative, and receives its name from the case of a potential with a single quartic self-interaction. Assuming the field only depends on the $\tau$ coordinate,
\begin{align}
	\frac{\delta V_{\rm 1-loop}(\varphi)}{\delta\varphi(\tau)} =
	\frac{\hbar}{2}\frac{1}{a^3(\tau)} \sum_{\ell=2}(\ell+1)^2
	G_{b,\ell}(\tau,\tau) a^3(\tau) V'''(\varphi(\tau))
	\equiv \Pi(\varphi(\tau))
	\label{eq:tadpoleFunction}
\end{align}
where Eq.~\eqref{eq:effPotOneLoop} and the specific form of the operator, Eq.~\eqref{eq:reducedFluctuationOp}, were used. The quantum corrected equation of motion can be written as
\begin{align}
	\left(-\frac{1}{\sqrt{g_E}}\partial_{i}\sqrt{g_{E}}\,g_{E}^{ij}\partial_{j}\varphi
	 + V'(\varphi(\tau)) + \Pi(\varphi(\tau)) \right) = 0.
\end{align}
If we consider splitting $\varphi$ into the bounce plus quantum corrections, using
the bounce equation, Eq.~\eqref{eq:bounceEqGeneral}, we have for the next to leading
order the differential equation
\begin{align}
	-\frac{1}{a^3(\tau)}\partial_{i}\sqrt{g_{E}}\,g_{E}^{ij}\partial_{j}\phi^{(1)}
 + V''(\phi^{(0)})\phi^{(1)} + \Pi(\phi^{(0)}(\tau)) + \mathcal{O}(\phi^{(1)\,2}) =
 0.
\end{align}
Here we can recognize the fluctuation operator of
Eq.~\eqref{eq:fluctuationsOpGeneral}, so that knowledge of the associated Green's
function, $\mathcal{G}(x,x')$, allows us to quickly compute the the quantum
corrections to the background through
\begin{align}
	\phi^{(1)}(x) &= -\int \d x' \mathcal{G}(x,x')a^3(\tau)\,\Pi(\phi^{(0)}(\tau))\\
	& =	-\frac{1}{2\pi^2}
	\sum_{\ell=0}(\ell+1)\int\d\Omega'_3 \d\tau' U_\ell ({\bf e}_x\cdot {\bf
	e}_{x'})G_\ell(\tau,\tau')\,\Pi(\phi^{(0)}(\tau')).
\end{align}
Here we have inserted our expression for the Green's function from
Eq.~\eqref{eq:fullGreenSolution}. Computing the integral over the solid angle gives
the following:
\begin{align}
	\phi^{(1)}(x) &= -
	\sum_{\ell=0, {\rm even}}(\ell+1) \int
	\d\tau'G_\ell(\tau,\tau')\, \Pi(\phi^{(0)}(\tau')),
\end{align}
where the integration over the Chebyshev polynomials is equal to $\pi$ for even
values of $\ell$ and $0$ otherwise. Moreover if the fluctuations are
assumed to occur only in the $\tau$-direction the only modes that are needed are the ones in the $\ell=0$ sector, thus
\begin{align}
	\phi^{(1)}(x) &= -\int\d\tau' G_0(\tau,\tau')\, \Pi(\phi^{(0)}(\tau')).
	\label{eq:quantumCorrBounce}
\end{align}
Once the quantum corrections are known, it is in principle viable to self-consistently evaluate the the one-loop effective action on $\phi^{(0+1)}$ to include specific two-loop effects such as dumbbell diagrams (see \cite{Ai:2018guc}).

\section{Numerical implementation and results}

We apply the techniques described in the previous sections to a case study for a given set of values for the parameters appearing in the rescaled potential Eq.~\eqref{eq:rescaledPot} where in addition we parametrize the value at the top of the potential as
\begin{align}
	V_0 = \frac{H^2 v^2}{\epsilon^2}.
\end{align}
The set of Mathematica notebooks and codes can be found in the \verb|false-vacuum-decay-desitter| GitHub repository  \footnote{\href{https://github.com/Stephan-Brandt/false-vacuum-decay-desitter}{\faGithub https://github.com/Stephan-Brandt/false-vacuum-decay-desitter}}. Specifically and throughout this section, we consider the same parameter set as the one considered in  \cite{Dunne2006}, namely
\begin{align}
	\beta = 45,\quad b= 0.25,\quad \epsilon=0.046\quad\text{and}\quad H=1,
	\label{eq:paramSet}
\end{align}
which allow for a direct comparison with their work, but does not allow to perturbatively include the effects of the bounce's quantum corrections self-consistently.

We first compute the bounce configuration solving Eq.~\eqref{eq:eomBounceRescaled}, demanding that $\phi(0)=\phi_+$ and $\phi(\pi/H)=\phi_-$. A solution satisfying these conditions exists exactly in the non-thin-wall case, due to the change of sign in the friction term, so that the starting point is extremely close to the true vacuum.

\begin{figure}[htbp]
	\centering
	\includegraphics[clip,width=.75\textwidth]{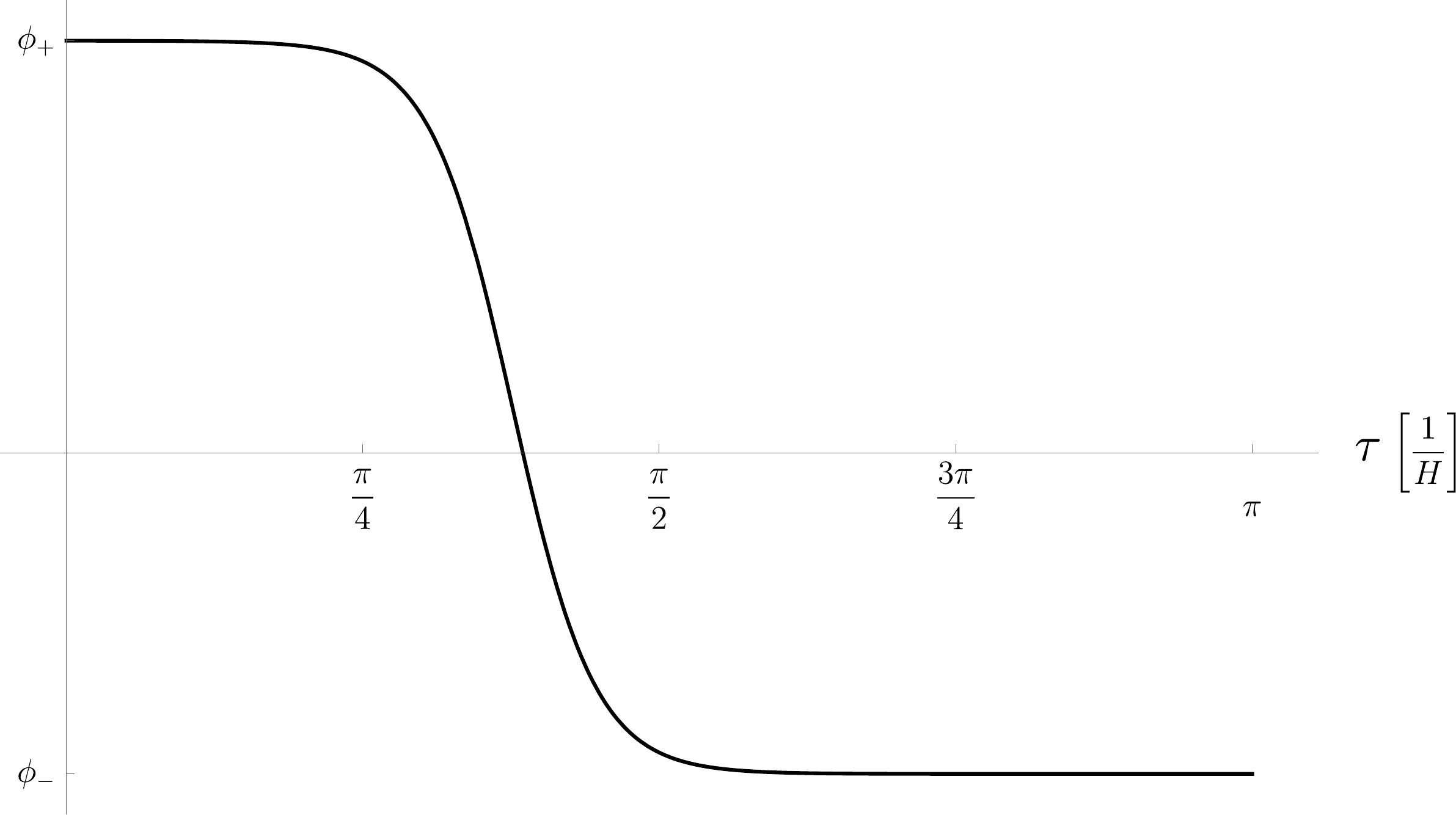}
	\caption{Profile of the classical background that interpolates between the local minima of the potential, $\phi_-$ and $\phi_+$.}
	\label{fig:bounce}
\end{figure}

Numerically, the problem is solved via a so-called shooting method, which turns the boundary value problem into a set of initial value problems. We keep the initial speed fixed to $0$ and vary the initial position of the candidate solution such that the method converges to the bounce by demanding the right behavior at $\sigma \rightarrow H \pi$, namely ensuring that the bounce approaches some value close to $\phi_-$ for $\sigma = H\pi - \varepsilon$. In order to find the initial position in the shooting method, we must update the initial position as follows. The algorithm looks for the minimum of the field configuration, and if it lies outside the range between true and false vacuum, the solution is said to have overshot, consequently the updated initial position is shifted further towards $\phi_-$. Alternatively, if the first local minimum lies between the two vacua, the solution undershot and more potential energy is required to reach the other vacuum, meaning the starting position is shifted closer to $\phi_+$. Iterating this procedure, the output converges to a Coleman-de Luccia-like bounce. An example of the profile of the bounce configuration used in the present study can be found in Fig.~\ref{fig:bounce}.
\label{sec:numericalImpl}

\begin{figure}[htbp]
	\centering
	\includegraphics[clip,trim=1.7cm 0 0 0, width=.65\textwidth]{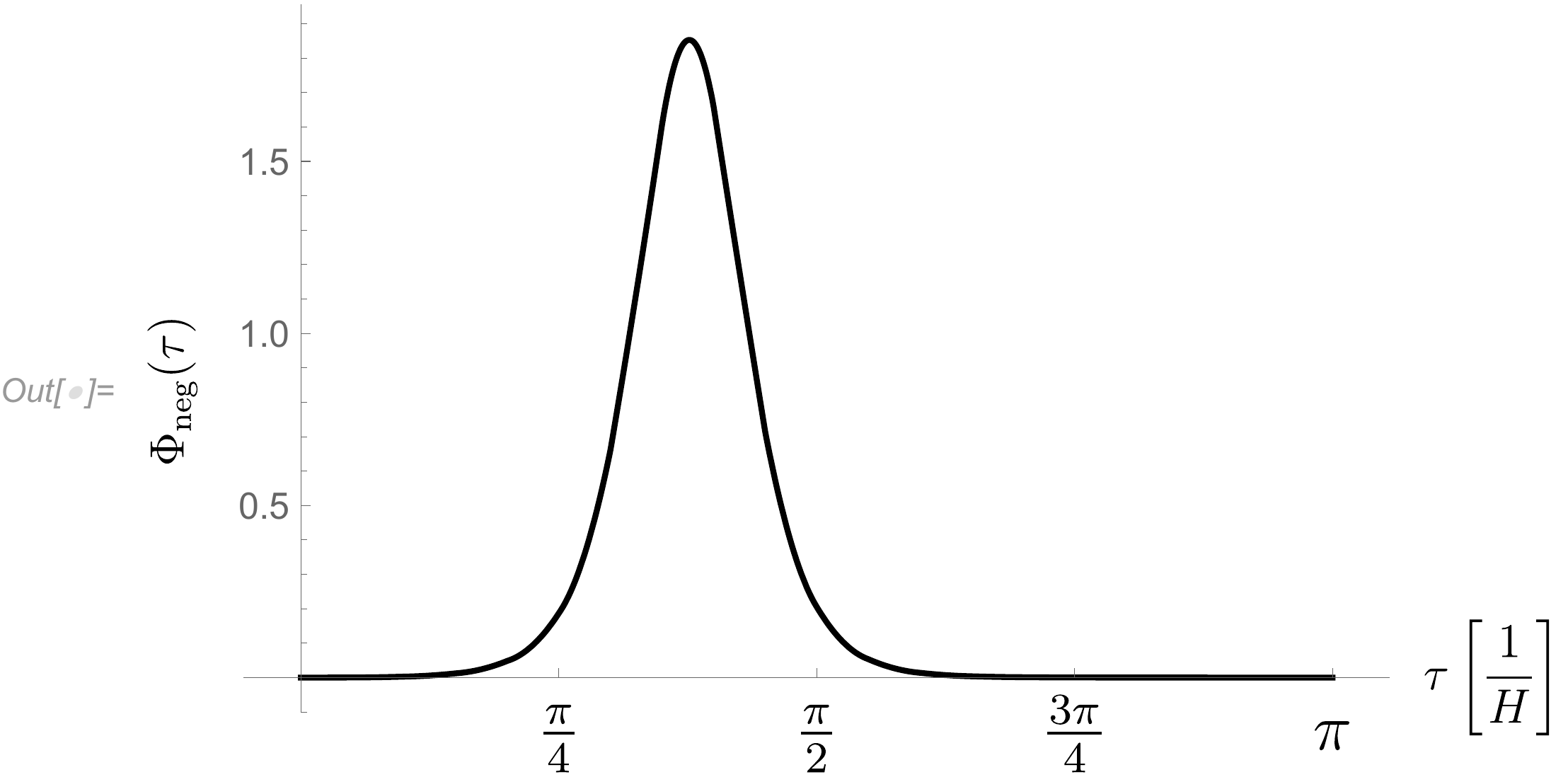}
	\caption{The Eigenfunction belonging to the negative eigenvalue $\lambda=-2.71191$ for the model with parameters as in Eq.~\eqref{eq:paramSet}.}
	\label{fig:negmode}
\end{figure}

Before solving for the full one-loop contributions, we extract the negative mode, essential in describing the tunneling process. This can be done straightforwardly using, for example, Mathematica's algorithm \verb|Eigensystem|. The negative mode was determined to be $\lambda=-2.71191$ for the original system. The corresponding eigenfunction is shown in figure \ref{fig:negmode}.

We search for possible zero modes numerically, for the set of parameters in Eq.~\eqref{eq:paramSet}. It should be noted, that an expansion in hyperspherical harmonics for the to-be fluctuations implies that if zero modes exist, they must come from the $\tau$-dependent factor. We find that no zero modes associated with the lower values of $\ell$ are present. That is, Eq.~\eqref{eq:jacobi} does not seem to have zero modes for $\ell=0$ or $\ell=1$. Moreover, we observe that the absolute value of the eigenvalues keeps increasing in every $\ell$ sector, which seems to indicate no zero modes will at all be present. We do not find any such modes, which, together with the expressions given in Subsec.~\ref{subsec:zeromodes}, indicate that the question of finding the zero modes in this scenario is subtle and requires a study of its own.

Having a numerical expression for the bounce configuration, we now solve for the one-loop contributions to the action, making no approximations regarding the scalar background.

\begin{figure}[htbp]
	\includegraphics[
	width=.85\textwidth]{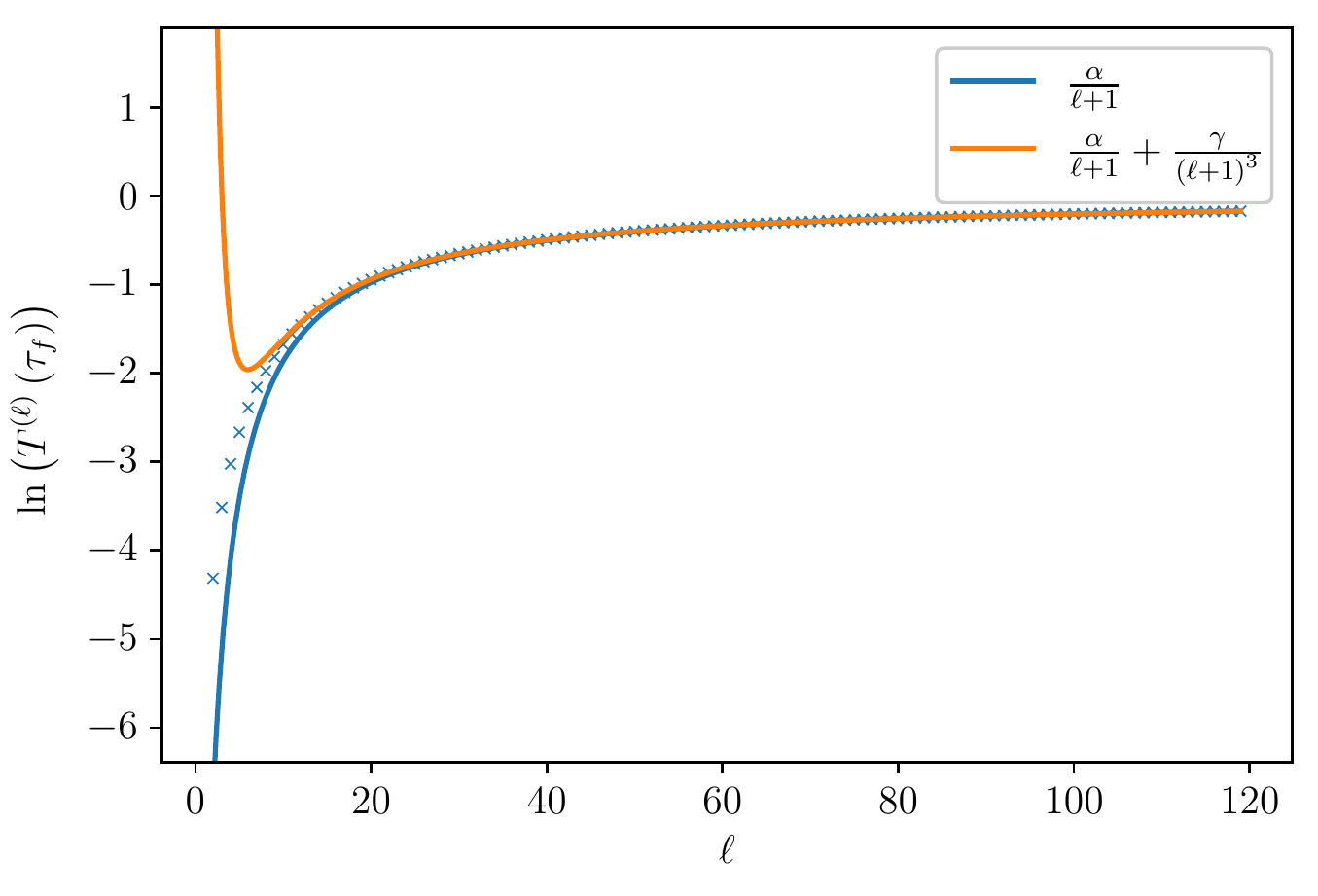}
	\caption{The individual crosses show the partial determinants $\ln T^{(\ell)}$, where $\alpha= -20.45768 $ and $\gamma= 330.41108$ as obtained through Eqs.\eqref{eq:wkbAlpha} and \eqref{eq:wkbGamma} for the set of parameters in Eq.~\eqref{eq:paramSet}. The enveloping functions follow the diverging behavior for large-$\ell$ that has to be subtracted.}
	\label{fig:partialdets}
\end{figure}

First, we discuss the Gel'fand-Yaglom method. As explained in Sec.~\ref{sec:funcDets}, we must solve the Jacobi equation, Eq.~\eqref{eq:ratio-operator}, and compute the specific ratio of its eigenfunctions as in Eq.~\eqref{eq:definitionT} and evaluate at $\tau_f=\pi$. With Mathematica, the analytical solution for the free case, written in Eq.~\eqref{eq:false-vacuum-fluctuation}, cannot be evaluated reliably at arbitrarily small times or times close to the Hubble horizon. To mitigate this problem, the function is tabulated for several intervals and patched together piecewise to generate the full solution. The value of the ratio-operator at time $\tau \rightarrow H\pi$ is sensitive to the initial time $\tau_{0} \ll 1$ from where we start to solve equation \eqref{eq:ratio-operator}. We solve the differential equation, determine $T(H\tau)$ for a sequence of initial times and subsequently extrapolate back to $\tau_{0} \rightarrow 0$. A plot of the partial determinants is shown in Fig.~\ref{fig:partialdets} and is seen to be in excellent agreement with  \cite{Dunne2006}.
We write down the coefficients of the divergent contributions with $\alpha$ and $\gamma$ respectively as found using the WKB approximation (Eqs.~\eqref{eq:wkbAlpha},\eqref{eq:wkbGamma})

\begin{figure}[htbp]
	\includegraphics[width=.75\textwidth]{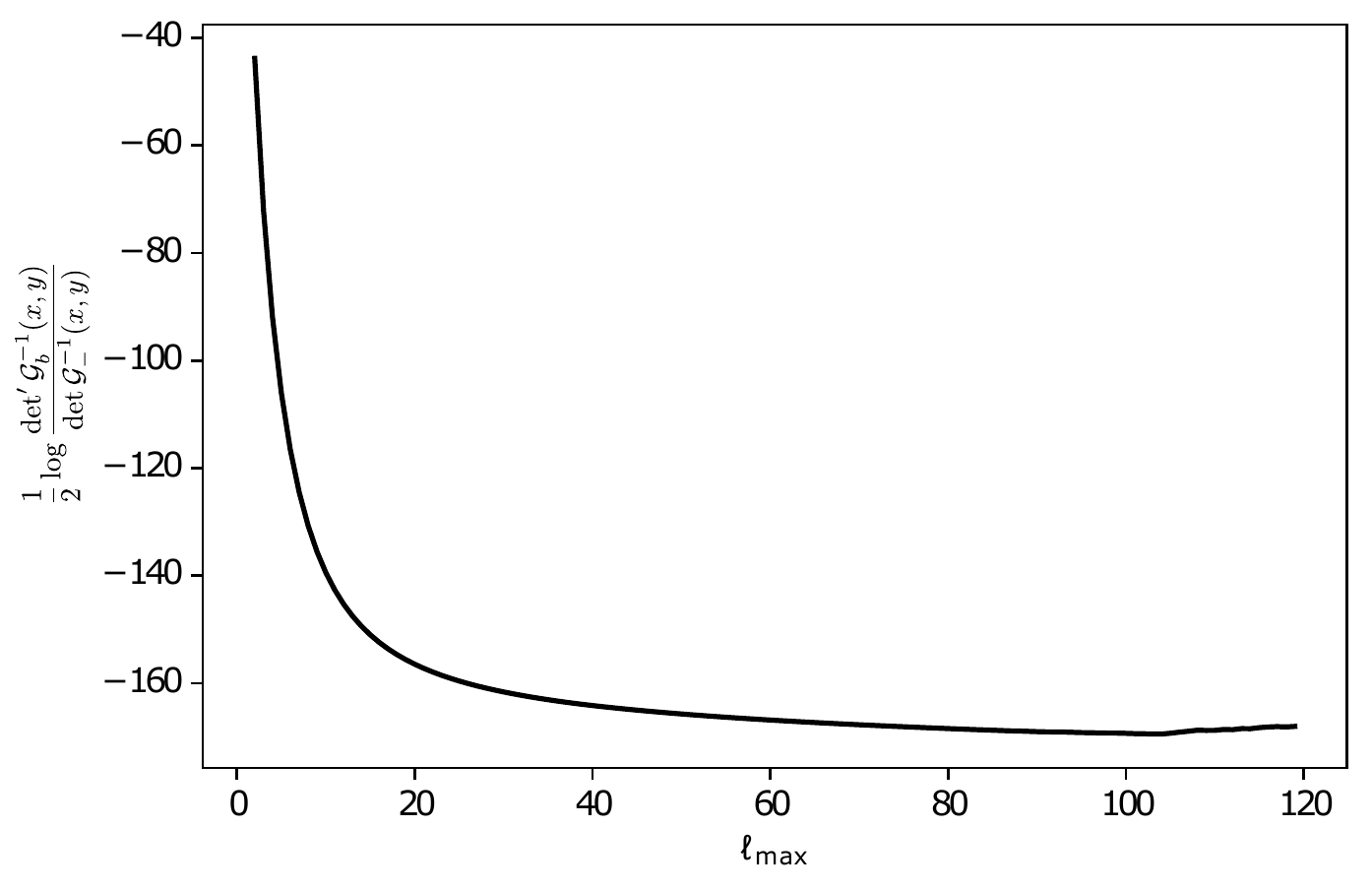}
	\caption{A plot of the series of partial determinants after regularizing the large-$\ell$ behavior. As we can see, there is rapid convergence in the middle section. However, at large values of $\ell$, the numerical noise begins to take over.}
	\label{fig:totdet}
\end{figure}

In order to check how accurately the WKB Ansatz solves the Jacobi equation, we check its remainder. Explicitly, we verify that the zeroth order WKB solution satisfies the differential equation well even for low values of $\ell$, where the spacetime dependence of the potential is expected to dominate, when compared to the $\ell$ term. The same is verified when including the next order WKB corrections. For large values of $\ell$, the $\ell$-term dominates and the corrections become extremely small, so the WKB approximates the exact solution. We regularize the one-loop expression, by removing the divergent terms, rendering the sum in Eq.~\eqref{eq:totaldet} finite. We plot the resulting series in Fig.~\ref{fig:totdet}.

A problem affecting the numerical precision is the poor quality of the numerical coefficients in Eq.~\eqref{eq:GelfandYaglom}, especially at small times. The aforementioned extrapolation scheme of $T(H\tau)$ is needed to carry out the computation in the range $50 < \ell < 100$ and enables us to estimate the full sum. For $\ell < 50$ the associated Legendre function can be evaluated for reasonably small times and there is no need for extrapolation. For $\ell > 120$, however, the minimal time for which we can evaluate the associated Legendre function becomes too large for a reliable extrapolation. Additionally, any deviation from the exact solution is multiplied by a factor $(\ell + 1)^2$. Thus, for $\ell > 100$ we observe the impact of the noisy behavior towards the lower right corner of Fig.~\ref{fig:totdet}.

\begin{figure}[htbp]
	\begin{center}
	\includegraphics[clip, trim=1.2cm 0cm 1.2cm 0, width=.7\textwidth]{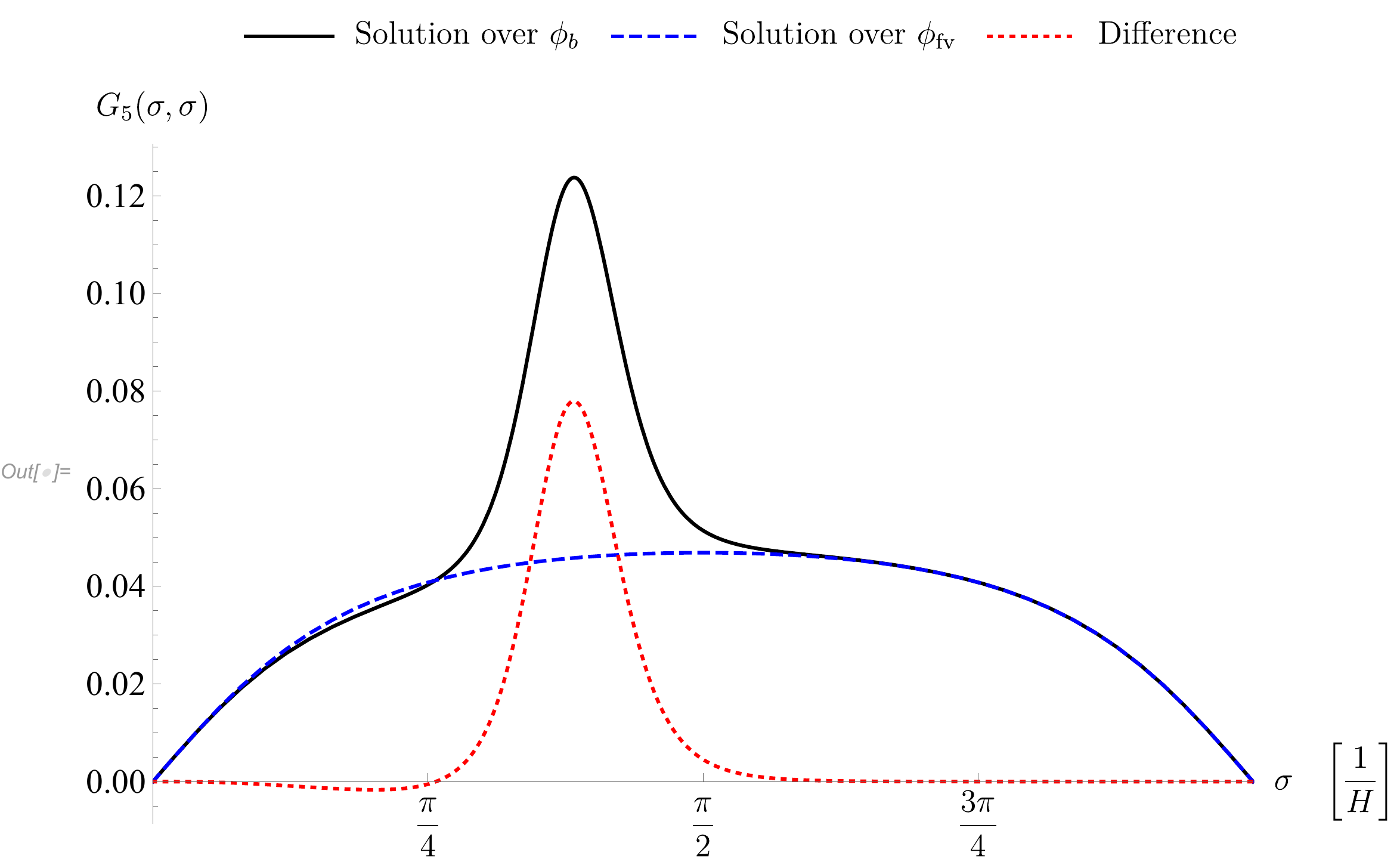}
	\end{center}
	\caption{Example of coincident Green's function which solves Eq.~\eqref{eq:reducedFluctuationOp} for $\ell=5$, plotted against $\sigma=H\tau$. For the parameters $\beta=45$, $b=1/4$, $\epsilon=0.046$ and $H=1$.}
	\label{fig:sampleCoinGreen}
\end{figure}

\begin{figure}[htbp]
	\begin{subfigure}{.44\textwidth}
		\includegraphics[width=\textwidth, trim=1cm 0 1cm 0]{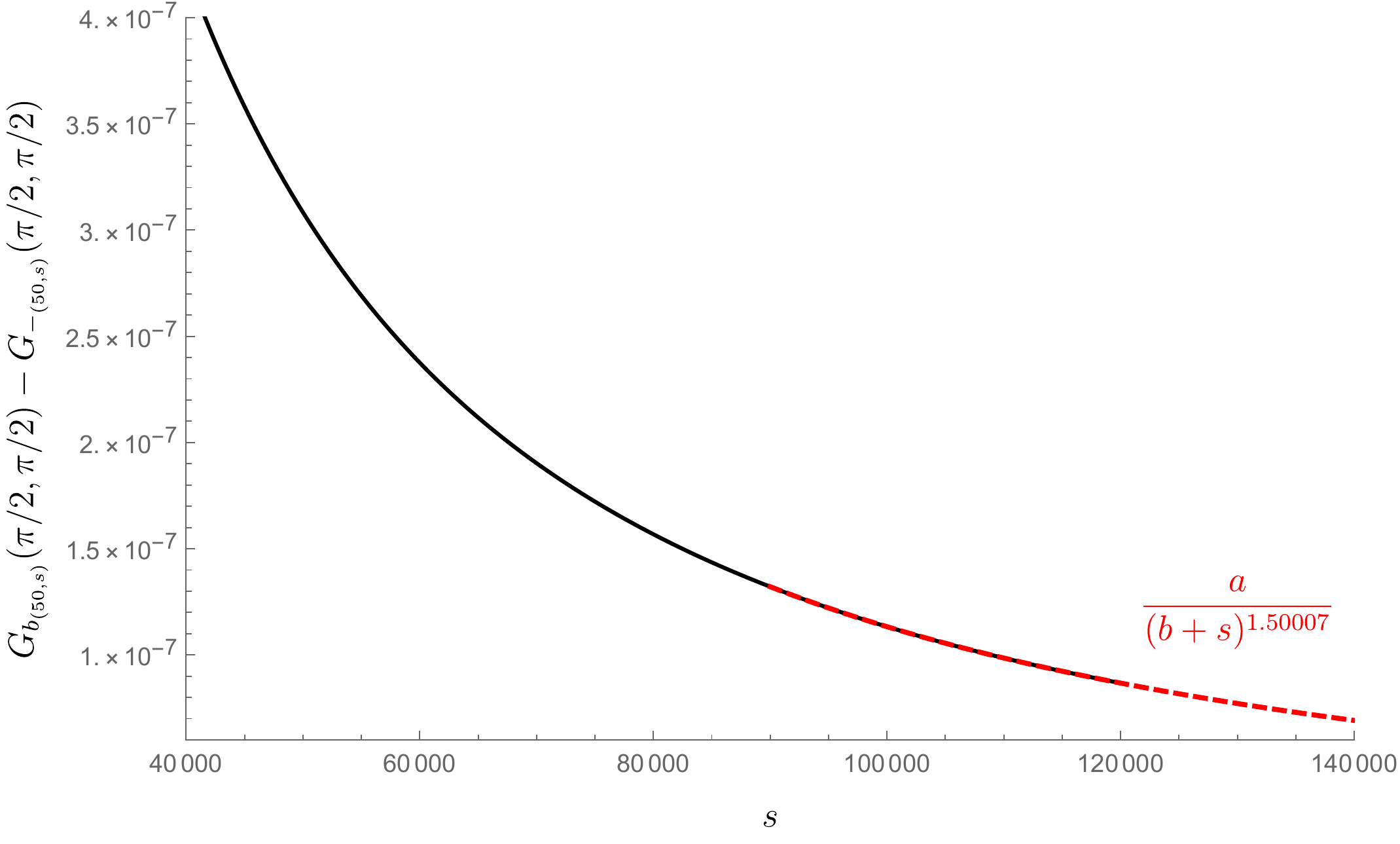}
		\caption{Example of the difference between coincident Green's functions at $H \tau=\pi/2$ and $\ell=50$, as a function of the deformation parameter $s$, together with a verification of its correct asymptotic behavior for large $s$.}
	\end{subfigure}
	\hspace*{.6cm}
	\begin{subfigure}{.44\textwidth}
		\includegraphics[width=\textwidth, trim=1cm 0 1cm 0]{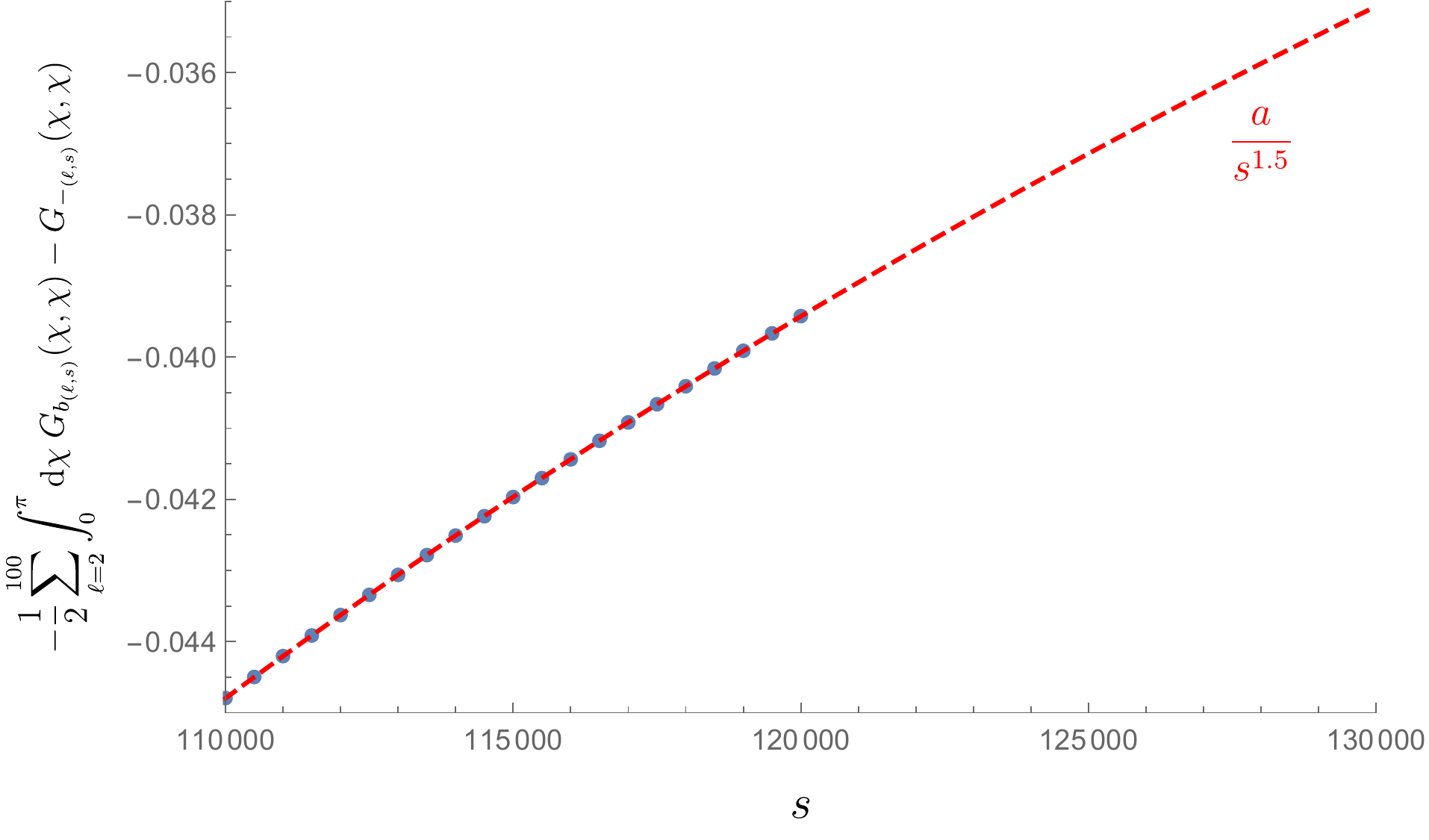}
		\caption{Sum of the integrated difference of Green's functions at the bounce and the false vacuum, respectively, up to $\ell=100$, depicting the extrapolation on the $s$-parameter corresponding to flat space.}
	\end{subfigure}
	\caption{Extrapolations for larger $s$ values for the difference and the integrated difference of the Green's functions evaluated at the bounce and the false vacuum respectively.}
	\label{fig:extrapolationsLegDet}
\end{figure}

In order to compute the one-loop contribution to the action via the Green's function method and include the gradients of the bounce exactly, we employ the decomposition shown in Eq.~\eqref{eq:ansatzReducedGreenFunctions} and solve the alternative system, Eq.~\eqref{eq:reducedFluctuationOp}. An example of the Green's functions obtained for a fixed value of $\ell$, together with the Green's function for the alternative fluctuation operator over the false vacuum, is depicted in Fig.~\ref{fig:sampleCoinGreen}. Especially for low values of $\ell$, it can be observed how the bounce configuration deforms the solution over the false vacuum in the region where gradients are strongest. For higher values, the Green's function over the bounce approaches the one over false vacuum (black and blue in the figure).
In order to compute the functional determinant, it is necessary to have available the Green's functions for the deformed operator in Eq.~\eqref{eq:deformedOp}. In practice, we use an analogous deformation but for the alternative operator and find the Green's functions numerically by scanning over $2\leq \ell\leq 100$ and $0\leq s\leq 120000$. The convergence of the integral in Eq.~\eqref{eq:logDetToGreens} in the $s$ direction is quite slow, and extrapolation of the Green's functions was used for higher values of the $s$ parameter once the behavior matched the power-law expected from the homogeneous expressions (see Fig.~\ref{fig:extrapolationsLegDet}).
For the computation of the Green's function, it was necessary to use a computation cluster and take advantage of the parallelizability of the problem, given that for each value of $s$ and $\ell$, the differential equation solver could be run independently. After collecting the output and including the extrapolation, it was possible to obtain the $\log\det$ one-loop contribution fully, including the full scalar background as desired. The different quantities computed are shown in Table~\ref{tab:results}. It is worth adding that the result coming from the Green's functions is particularly sensitive to the precision used in the computation. Initially, using a working precision of five digits, we obtained a result much closer to the Gel'fand-Yaglom result and it is only when increasing this precision that we can be certain of the result that includes three significant figures reported in the table. This points at the possibility of improving the Gel'fand-Yaglom method by for example taking an even finer partition for the numerical treatment of the Legendre functions in Eq.~\eqref{eq:false-vacuum-fluctuation}.
\begin{figure}[htbp]
	\centering
	\includegraphics[clip,trim=1.9cm 0cm 0cm 0,width=.75\textwidth]{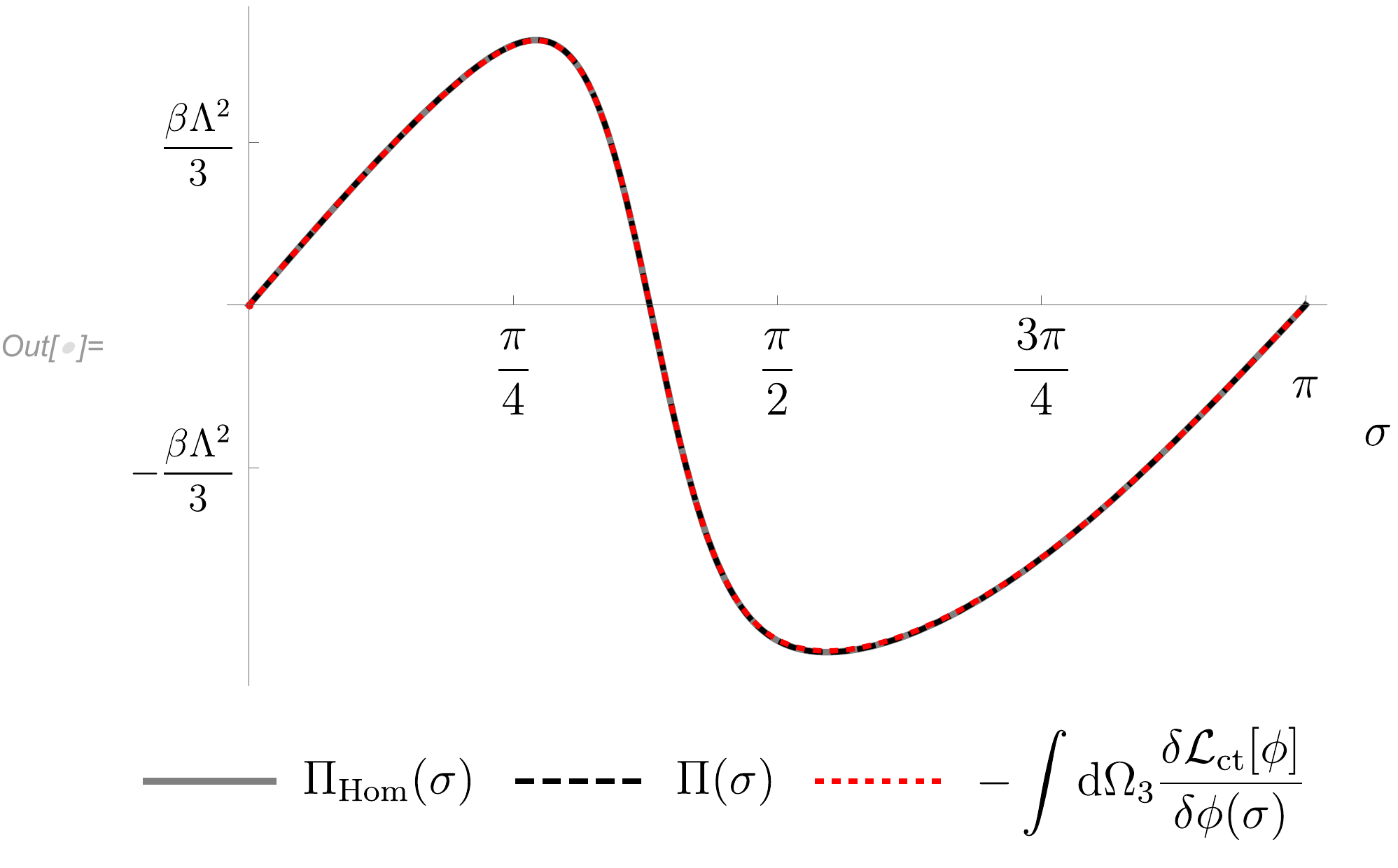}
	\caption{Different tadpole functions plotted against the time coordinate
		$\sigma= H\tau$. The bare homogeneous tadpole corresponding to the tadpole function obtained with the function in Eq.~\eqref{eq:homCoincidentGF}, is shown as a continuous gray line, the bare exact tadpole computed numerically is shown as dashed lines and
		the negative functional derivative of the counter-terms capturing the divergent
		pieces is drawn as a red dotted curve. The upper cutoff for $\ell$ was chosen $\Lambda>200$ for this plot and the parameters in \eqref{eq:paramSet}.}
	\label{fig:tadpoleBareAndDCTs}
\end{figure}

\begin{figure}[!htbp]
	\includegraphics[clip,trim=0cm 0 0 0, width=.45\textwidth]{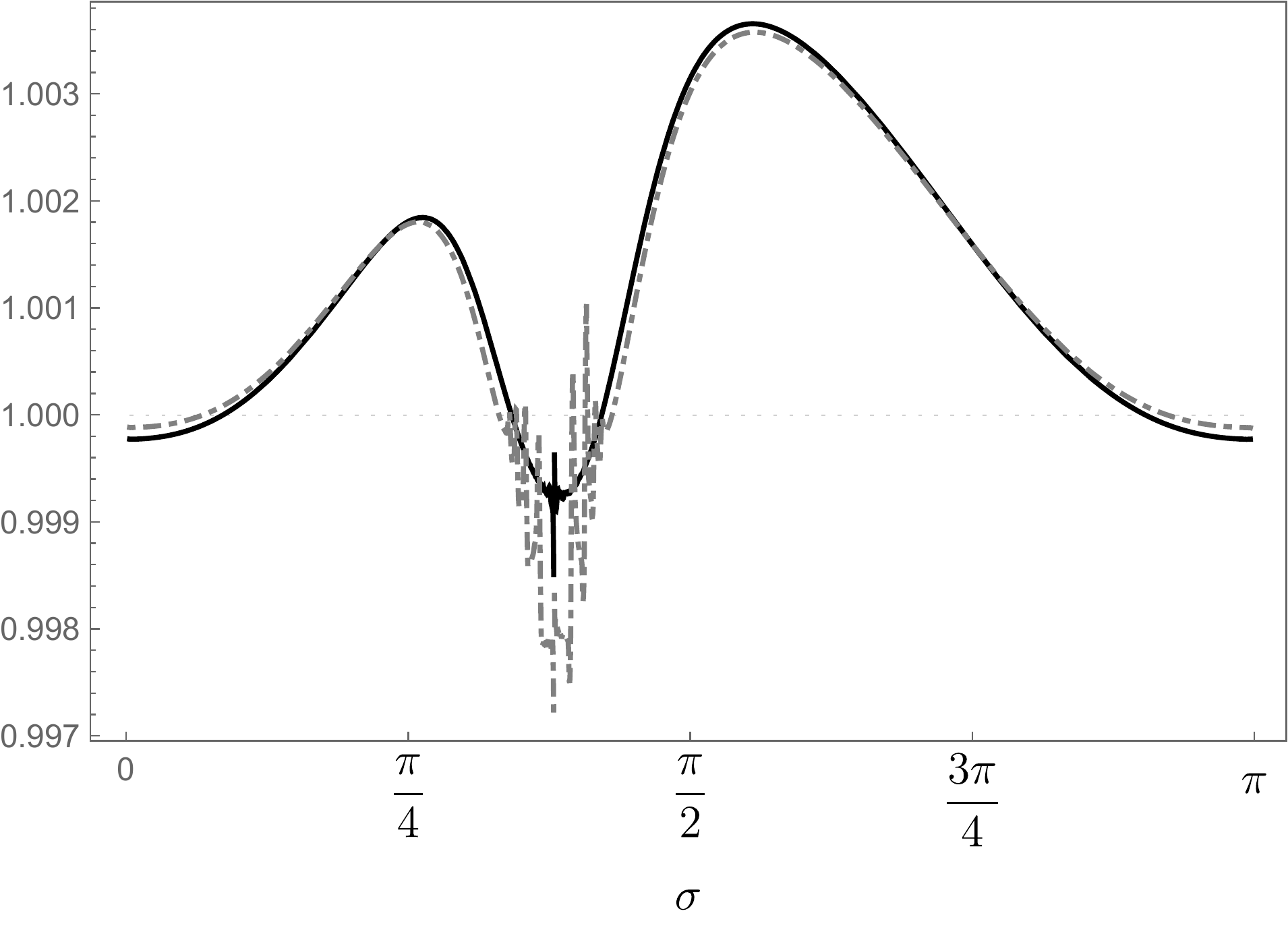}
	\quad
	\includegraphics[clip,trim=0cm 0 0cm 0cm,width=0.45\textwidth]{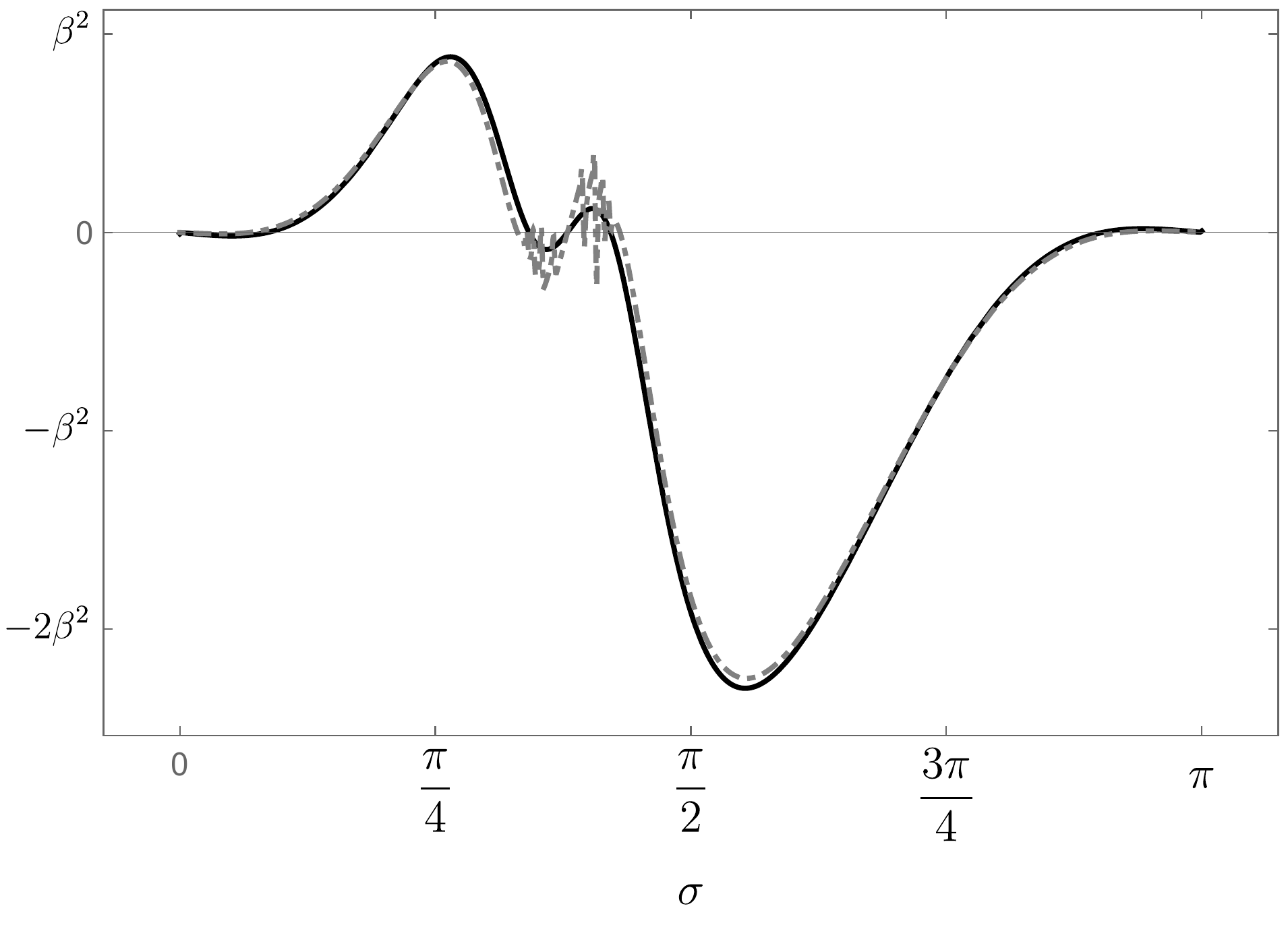}
	\captionsetup{justification=raggedright}
	\caption{(Left) Plot depicting the ratio between the homogeneous tadpole function and the negative of the	divergent pieces computed via WKB(dashed gray line), showing tachyonic regions between $\pi/4$ and $\pi/2$. Also the ratio between the exact (numerical) tadpole function and the negative divergent pieces computed via WKB (solid black curve). (Right) Renormalized homogeneous tadpole function (dashed gray) and renormalized numerical tadpole (solid black) for the benchmark parameters in Eq.~\eqref{eq:paramSet} and against $\sigma=H\tau$.}
	\label{fig:tadpoleRatioAndrenormTadpole}
\end{figure}

\begin{figure}[!htbp]
	\begin{flushleft}
	\hspace*{2.5cm}
	\includegraphics[clip,trim=1.9cm 0 0 0, width=.6\textwidth]{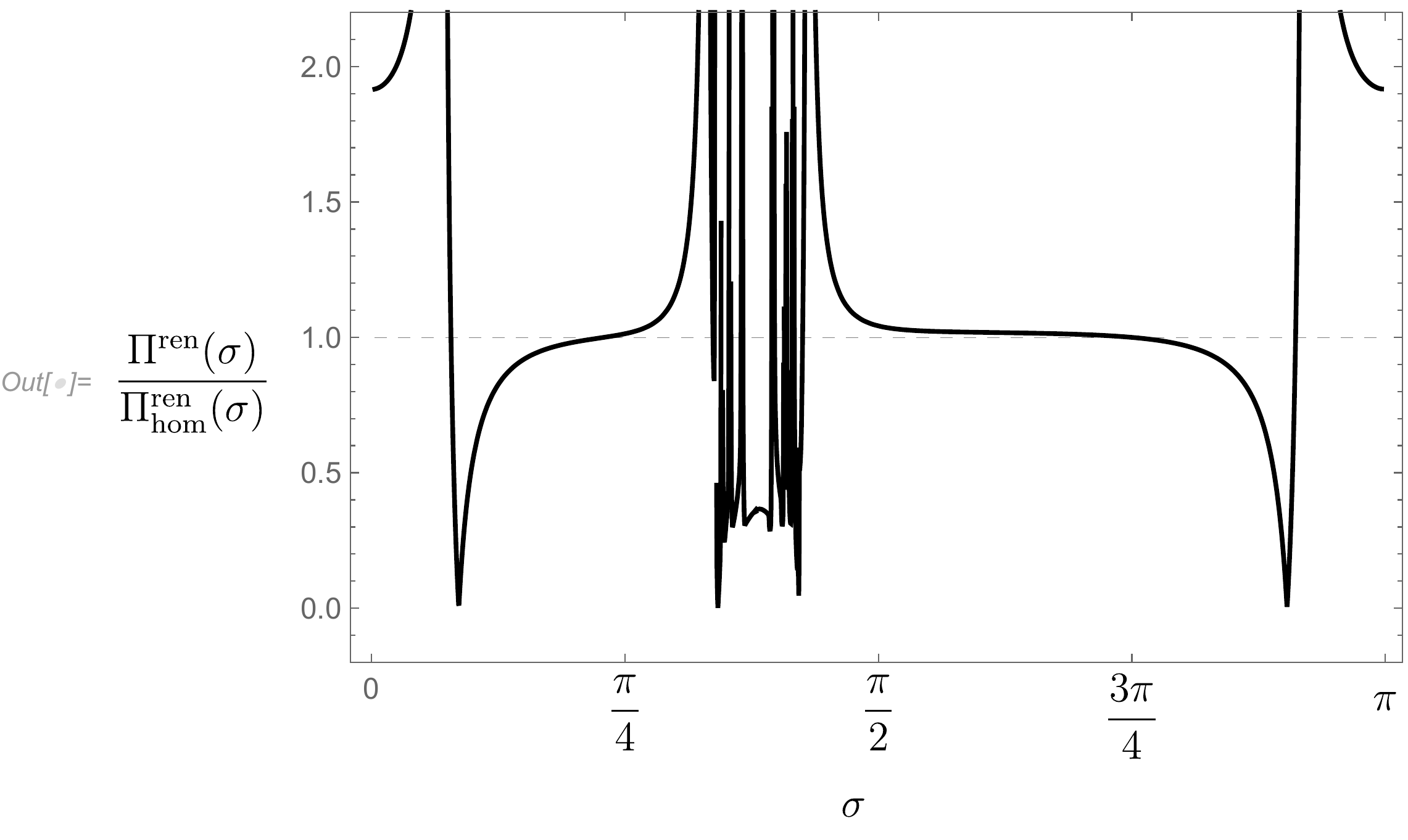}
	\caption{Ratio between the renormalized homogeneous tadpole function and the renormalized numerical tadpole function. The spikes towards the edges come from the non-smooth extension of the bounce profile and are not physical effects.}
	\label{fig:ratioRenormTadpoles}
	\end{flushleft}
\end{figure}

As proof of concept we further compute the tadpole function shown in Eq.~\eqref{eq:tadpoleFunction} and the quantum corrections to the bounce, Eq.~\eqref{eq:quantumCorrBounce}. For this purpose, it is necessary to renormalize the theory and use analytical expressions for the divergent pieces, only then we can obtain a renormalized tadpole function. In Figure \ref{fig:tadpoleBareAndDCTs}, we compare the bare tadpole function with the corresponding divergent pieces coming from the renormalization procedure. We find excellent agreement between the two for the cutoff parameter ($\Lambda>100$), meaning the WKB homogeneous Green's functions capture the UV divergences well. The ratio of these functions can be found in the left plot of Figure~\ref{fig:tadpoleRatioAndrenormTadpole}, where it can be compared with the ratio of its homogeneous analogue. Once the divergent pieces are removed, as in an MS-like scheme, we obtain the plot on the right in Figure~\ref{fig:tadpoleRatioAndrenormTadpole}, where we observe that oscillations are typically of the size of the mass scale of the field, $\beta H^2$. We also compare the renormalized homogeneous tadpole with the renormalized exact tadpole, in order to isolate the gradient effects, see Fig.~\ref{fig:ratioRenormTadpoles}

\begin{figure}[!htbp]
	\includegraphics[clip,trim=0cm 0 0cm 0,width=.7\textwidth]{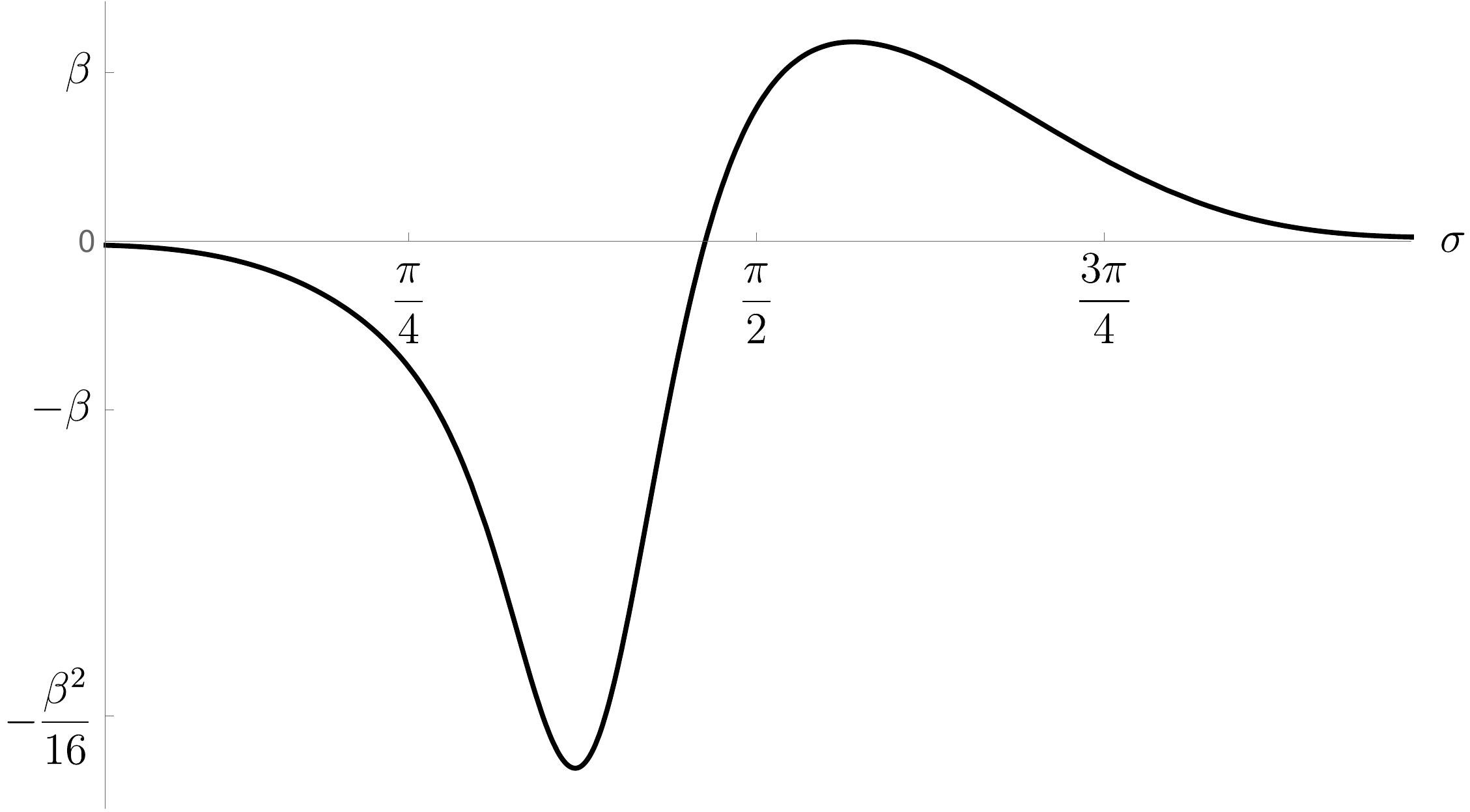}
	\caption{Quantum corrections for the bounce solution for the set of benchmark
	parameters chosen against $\sigma=H\tau$.}
	\label{fig:quantumCorrectionsBounce}
\end{figure}

Once the tadpole function is computed and renormalized, it is possible to compute the quantum corrections to the background Eq.~\eqref{eq:quantumCorrBounce}. For our set of parameters, we fall out of the perturbative regime, and the corrections cannot be included at face value. Nonetheless, we  conclude that the quantum corrections come out proportional to the finite parts of the tadpole function since the $\ell=0$ sector has a Green's function which is of order one. Thus, the bounce corrections are of order $\beta^2$ as can be seen in Figure \ref{fig:quantumCorrectionsBounce}.
\begin{table}[htbp]
	\centering
	\renewcommand*{\arraystretch}{1.3}
	\begin{tabular}{|c||c|c|c|c|c|c|}
		\hline
		Contribution & $S_0$ & $B^{(1)\,\rm  hom}$ & $B^{(1)\,\rm GY}$ & $B^{(1)\,\rm GF}$ & $\%^{\rm GY}$ & $\%^{\rm GF}$ \\
		\hline
		\hline
		Value & 620 & -169 & -164 & -158 & 3.0 & 6.5\\
		\hline
	\end{tabular}
	\caption{Summary of the different contributions to the effective action for the benchmark point.}
	\label{tab:results}
\end{table}

\section{Conclusions}
\label{sec:conclusions}

We perform the computation of the one-loop effective action over an inhomogeneous scalar background representing tunneling phenomena between two phases of a scalar field theory. We consider a fixed de Sitter geometry together with a potential displaying two non-degenerate minima, which allows for transitions from one to the other. For the computation of the effective action, we lift the common assumption of a thin wall and include the effects of the gradients of the scalar field at the level of the one-loop corrections. For a generic benchmark point, we conclude that considering the bounce background, the one-loop effects are corrected by about 3-6\% when compared to the homogeneous approximation of the one-loop effects, depending on the method and its precision.

We perform the numerical computation of the full one-loop effects, i.e. homogeneous plus gradients, employing two available methods in the literature and find them consistent with each other. First, we employ the Gelfand-Yaglom method for the computation of the one-loop effects. Second, we employ the Green's function method using resolvents. Both methods are implemented in completely independent ways and give compatible results for our benchmark computation up to the percent level when compared with each other.

We highlight the advantage of the Green's function method over the Gelfand-Yaglom one, where the former provides simultaneously two-point correlation functions and allows for the computation of higher-order effects, as was indicated in this document. For potentials that allow for a perturbative treatment in the couplings, we see no obstacle in including the quantum corrections to the scalar background in a consistent way.

We carry out the full computation of the one-loop effects and the regularization of this model, completing some of the open ends in Dunne's previous study \cite{Dunne2006}. This is done via the identification of divergent terms in a WKB expansion of the ratio of Jacobi equations or the fluctuations operator. We give explicit expressions for the homogeneous estimates to the tadpole functions and thus to the quantum corrections to the scalar background. However, we do not address the task of interpreting the divergent pieces as counterterms of a local theory. Although it is known  \cite{Birrell:1982ix}, that the specific case of de Sitter space can be renormalized in that sense, we leave this matter for future considerations.

After the renormalization procedure, we obtain the tadpole functions and thereafter find quantum corrections to the bounce configuration. Owing to the size of the $\beta$ parameter, we cannot include the size of the corrections directly. Nonetheless, by observing the leading contributions of the renormalized homogeneous expressions, we can see how the corrections appear to be proportional to $\beta^2$ and completely in line with previously related studies where the relative corrections to the bounce are found to be below 10\%.

There are several new questions to be answered. There is no clear picture of what happens to the zero modes expected in the $\ell=1$ sector and how the lack of them would impact the interpretation of the tunneling process as happening due to the appearance of bubbles. Instead, we can only speak of an overall probability for the field to transition globally from one vacuum to the other. It is known that multi-bounce solutions also exist  \cite{Hackworth:2004xb} and contribute to the action, and it would be interesting to consider their contributions compared to the gradients. The impact of the curvature scale $H^{-1}$ is expected to be more and more relevant as it tends toward the length determined by the mass scale of the field, but its exact effect on the relevance of gradients and the overall decay rate requires further examination.

\section*{Acknowledgments}
We would like to thank Björn Garbrecht, Martin S. Sloth and Carlos Tamarit for contributing with valuable comments and insight on this project.
Further thanks go to Gordian Edenhofer and Lode Pollet for advice and assistance concerning performance improvements in the numerical implementation.
Most of the numerical computation was performed using the computational cluster of the physics department at the Technical University of Munich. This work was supported by the Research Fund Denmark grant 0135-00378B.

\bibliography{references.bib}

\end{document}